# Transiting Exoplanet Atmospheres in the Era of JWST

Eliza M.-R. Kempton[1] and Heather A. Knutson[2]

[1]*Department of Astronomy, University of Maryland, College Park, MD 20742, USA*

[2]*Division of Geological and Planetary Sciences, California Institute of Technology, Pasadena, CA 91125, USA*

## ABSTRACT

The field of exoplanet atmospheric characterization has recently made considerable advances with the advent of high-resolution spectroscopy from large ground-based telescopes and the commissioning of the James Webb Space Telescope (JWST). We have entered an era in which atmospheric compositions, aerosol properties, thermal structures, mass loss, and three-dimensional effects can be reliably constrained. While the challenges of remote sensing techniques imply that individual exoplanet atmospheres will likely never be characterized to the degree of detail that is possible for solar system bodies, exoplanets present an exciting opportunity to characterize a diverse array of worlds with properties that are not represented in our solar system. This review article summarizes the current state of exoplanet atmospheric studies for transiting planets. We focus on how observational results inform our understanding of exoplanet properties and ultimately address broad questions about planetary formation, evolution, and diversity. This review is meant to provide an overview of the exoplanet atmospheres field for planetary- and geo-scientists without astronomy backgrounds, and exoplanet specialists, alike. We give special attention to the first year of JWST data and recent results in high-resolution spectroscopy that have not been summarized by previous review articles.

# 1. INTRODUCTION

## 1.1. *A Historical Perspective*

As soon as the first exoplanets were discovered in the 1990s, the quest to characterize these unique objects in more detail began in earnest. Images from science fiction movies come to mind when we picture alien planets, but these discoveries provided us with an exciting new opportunity to actually *measure* the atmospheric properties of extrasolar worlds. The first transiting exoplanet was discovered in 2000 (Charbonneau et al. 2000). These are planets with orbits that track directly in front of their host stars as viewed by an Earthbound observer, producing a small dip in the amount of light received, known as a transit. We focus on transiting planets in this review article because they present special opportunities for measuring atmospheric properties. The clever techniques for transiting exoplanet atmospheric characterization that have been developed by the astronomical community (described in detail in Section 1.3) are all premised on using the known orbital geometry of the system to extract the planetary signal from the combined light of the planet and host star. These techniques have been applied with great success over the last two decades to measure a host of atmospheric properties.

The first detection of an exoplanet atmosphere occurred in 2002 (Charbonneau et al. 2002). The planet, HD 209458b, was the only known transiting planet at the time (although that was not the case for long), and it belongs to a broader class of exoplanets referred to as 'hot Jupiters'. Such planets are aptly named for their large sizes and small orbital separations — HD 209458b orbits its Sun-like host star every 3.5 days at an orbital distance of 0.05 AU, and it has a radius of 1.35 Jupiter radii ($R_J$). By measuring a small amount of excess absorption during transit at the wavelength of the sodium resonance doublet (589.3 nm) with the STIS instrument on the Hubble Space Telescope (HST), Charbonneau et al. (2002) inferred the presence of gaseous sodium in the planet's atmosphere. Although subsequent studies of HD 209458b's sodium absorption signal with ground-based high resolution spectrographs have revealed that this measurement may be biased by deformations in the stellar line shape due to the planetary transit (Casasayas-Barris et al. 2020, 2021), this first atmospheric measurement unquestionably marked the birth of a new field of exoplanet atmospheric characterization studies.

Not long after, came the first measurements of exoplanetary thermal emission via *secondary eclipse* (Deming et al. 2005; Charbonneau et al. 2005), which occurs when a planet passes *behind* its host star. Then, in 2007, the first phase curve observations of thermal emission versus orbital phase were obtained for the hot Jupiter HD 189733b (Knutson et al. 2007). The thermal emission measurements were all made with NASA's Spitzer Space Telescope, which became a workhorse for infrared (IR) characterization of exoplanet atmospheres before it was decommissioned in late 2020. Other important firsts include the measurement of escaping gas from an exoplanet atmosphere (VidalMadjar et al. 2003), and the first robust detections of molecules and (more tentatively) high-altitude winds using a novel cross-correlation spectroscopy technique with high-resolution spectrographs on ground-based telescopes (Snellen et al. 2010). More details on all of these observational techniques can be found in Section 1.3. All of the aforementioned observations were of hot Jupiter targets. The first atmospheric spectrum of an object smaller than Neptune was obtained in 2010 for the planet GJ 1214b (Bean et al. 2010), ultimately indicating the presence of a thick layer of clouds or haze (Kreidberg et al. 2014a). In 2018 the first thermal emission measurement was made for a rocky, terrestrial exoplanet, LHS 3844b, disappointingly indicating the lack of any atmosphere at all (Kreidberg et al. 2019).

Today, exoplanet atmospheric characterization has become its own *bona fide* sub-field of astronomy. Detections of several dozen atomic and molecular species along with clouds and hazes have been claimed in the literature for over 100 individual exoplanets[1] (see e.g. Burrows 2014; Madhusudhan et al. 2016; Deming & Seager 2017; Madhusudhan 2019). We note that some of these detections have been made at high statistical significance, whereas others are more tentative or ambiguous, so we encourage the casual reader of the exoplanet atmospheres literature to do so with a critical eye. All of these atmospheric characterization studies have been helped along by the discovery of thousands of transiting exoplanets [2] with ground-based (e.g. HAT, WASP, MEarth, Speculoos) and space-based (e.g. CoRoT, Kepler, TESS) surveys. On the population level, tantalizing hints of planetary diversity have been uncovered, and well-founded attempts are being made to tie statistical trends in atmospheric properties to underlying theories of planet formation and evolution (e.g. Sing et al. 2016; Tsiaras et al. 2018; Welbanks et al. 2019; Mansfield et al. 2021; Goyal et al. 2021; Changeat et al. 2022; Deming et al. 2023; Brande et al. 2023; Gandhi et al. 2023).

---

[1] At the time of publication, these two websites provide useful lists of published exoplanet atmospheric characterization results: http://research.iac.es/proyecto/exoatmospheres/index.php and https://exoplanetarchive.ipac.caltech.edu/cgi-bin/atmospheres/nph-firefly?atmospheres.

In late 2021, the James Webb Space Telescope (JWST) launched successfully, and scientific operations began in the summer of 2022. The telescope's large aperture and IR observing capabilities have opened the door to studies of smaller and colder planets than had previously been possible (e.g. Kempton et al. 2023; Greene et al. 2023; Zieba et al. 2023). The high signal-to-noise (S/N) spectra delivered by JWST for larger and hotter planets additionally enable analyses of processes that had remained hidden in earlier datasets such as inhomogeneous cloud formation (Feinstein et al. 2023) and photochemistry (Tsai et al. 2023). This new era of exoplanet characterization with JWST is accompanied by a windfall of ground-based exoplanet data using recently commissioned highresolution spectrographs (e.g. CARMENES, ESPRESSO, CRIRES+, IGRINS, GIANO, MAROONX, etc.) that are providing detailed compositional measurements for hot and ultra-hot giant planets (e.g. Birkby 2018; Giacobbe et al. 2021; Pelletier et al. 2023; Gandhi et al. 2023). The first JWST exoplanet observations have already been transformative, as have studies that have detected a slew of atomic, ionic, and molecular species in hot Jupiter atmospheres from the ground. These recent results will be summarized in the following sections along with the pre-existing context from two decades of exoplanet atmospheric characterization studies.

---

[2] A database that maintains a list of all known exoplanets: https://exoplanetarchive.ipac.caltech.edu/.

## 1.2. *Exoplanet Demographics*

We currently know of more than 10,000 extrasolar planets and planet candidates[3], most of which orbit stars with masses ranging from 0.5 – 1.5× the mass of the Sun (for astronomers, this corresponds to F through early M spectral types). If we exclude planets detected using the microlensing technique (a small fraction of this total), nearly all of these planets orbit stars in our local neighborhood[4] of the Milky Way galaxy. This means that when we discuss the properties of extrasolar planets in subsequent sections, we are implicitly focusing on planets orbiting relatively nearby and (unless specified otherwise) Sun-like stars. In this section, we provide a brief overview of this exoplanet population for the non-expert reader. We begin by briefly summarizing the two detection techniques most commonly used to find exoplanets and their corresponding sensitivities to different kinds of planets. For readers interested in learning more about complementary microlensing and direct imaging techniques, we recommend reviews by Gaudi (2022) and Currie et al. (2023). For a more comprehensive overview of exoplanet demographics, we recommend the review by Gaudi et al. (2021).

### 1.2.1. *Detection Techniques*

The first planet orbiting a Sun-like star was detected using the radial velocity technique (Mayor & Queloz 1995). This technique relies on the fact that a star and planet will orbit around their mutual center of mass. This causes the star's spectrum to be Doppler shifted as it moves towards and then away from the observer. The semi-amplitude of this Doppler shift is largest for massive planets with short orbital periods (e.g. Fischer et al. 2014); smaller planets on more distant orbits have smaller radial velocity semi-amplitudes and are correspondingly harder to detect[5]. By measuring a planet's radial velocity semi-amplitude, we can place constraints on its mass (technically $M_p\sin(i)$, where $M_p$ is the planet mass and $i$ is the orbital inclination), orbital period, and orbital eccentricity. We can then convert this orbital period to an orbital semi-major axis using Kepler's third law.

---

[3] For the latest numbers see Footnote 2 above. Most unconfirmed candidates were detected using transit surveys and it is likely that this sample contains some false positives, which are typically multiple star systems where one stellar component eclipses another. Transiting planet candidates can be validated statistically using the transit light curve shapes and other complementary information, such as adaptive optics imaging to resolve nearby stars (e.g., Morton et al. 2016; Giacalone et al. 2021), or they can be confirmed directly by radial velocity measurements of the planet masses.

[4] The distance from Earth to the center of the Milky Way is approximately 8.2 kpc (Bland-Hawthorn & Gerhard 2016), while most known exoplanets are located within a few hundred pc of the Earth's location (see https://exoplanetarchive. ipac.caltech.edu/).

Over the past decade the radial velocity technique has been overtaken by the transit technique, which is responsible for identifying most of the exoplanets known today. This technique focuses on planetary systems where the planet passes in front of its host star as seen from the Earth. During a transit, the planet will block part of the star's light. The amount of light blocked tells us the radius of the planet relative to that of the star, and the intervals between transits tell us the planet's orbital period. If we assume that the planet orbits are randomly oriented, the probability of seeing a transit P is given by P = $R_*/a$, where $R_*$ is the stellar radius and $a$ is the planet's orbital semi-major axis (Winn 2010). This means that transit surveys are biased towards close-in planets; this bias is even stronger than that of radial velocity surveys. Transit surveys also detect large planets more easily than small planets, as they block more of the star's light.

It is very challenging to detect Earth analogues orbiting Sun-like stars in current radial velocity and transit surveys. Fortunately, the size of both the transit and radial velocity signals increase with decreasing stellar mass. As a result, it is significantly easier to detect small planets orbiting small stars ('M dwarfs'). Small stars are also significantly less luminous than the Sun, and the orbital periods corresponding to Earth-like insolations are much closer in. This means that most small (approximately 1–2 R$_\oplus$) transiting planets that are amenable to atmospheric characterization orbit low-mass stars. This has important implications for our understanding of the population-level properties of small rocky exoplanets.

---

[5] The orbital motion of the Earth around the Sun produces a sinusoidal radial velocity signal with a semi-amplitude of 8.95 cm s$^{-1}$. We can use Equation 1 in Fischer et al. (2014) to calculate that a Jovian planet orbiting a Sun-like star with an orbital period of a few days would have a radial velocity semi-amplitude that is a factor of ~$10^3$ larger.

### 1.2.2. *Planet Types and Order-of-Magnitude Occurrence Rates*

As noted in Section 1.1, the close-in gas giant exoplanets known as 'hot Jupiters' were the first type of exoplanet detected in orbit around nearby Sun-like stars. These planets are relatively rare, with an order-of-magnitude occurrence rate of approximately 1% for Sun-like stars (e.g., Petigura et al. 2018; Dattilo et al. 2023). Gas giant planets at intermediate orbital distances (orbital periods of ∼ 10−100 days) are often referred to as 'warm Jupiters', and have a moderately enhanced occurrence rate relative to hot Jupiters (e.g., Fernandes et al. 2019; Fulton et al. 2021). Gas giant planets at larger separations (orbital periods greater than several hundred days) are typically referred to as 'cold Jupiters'. The most precise estimates of the occurrence rates of cold Jupiters currently come from radial velocity surveys, as there are very few transiting gas giant planets at these separations (e.g. Foreman-Mackey et al. 2016). These surveys indicate that the occurrence rate of gas giant planets rises dramatically as we move farther away from the star (e.g., Fernandes et al. 2019; Fulton et al. 2021), with 14 ± 2% of Sun-like stars hosting a gas giant planet between 2 − 8 AU (Fulton et al. 2021).

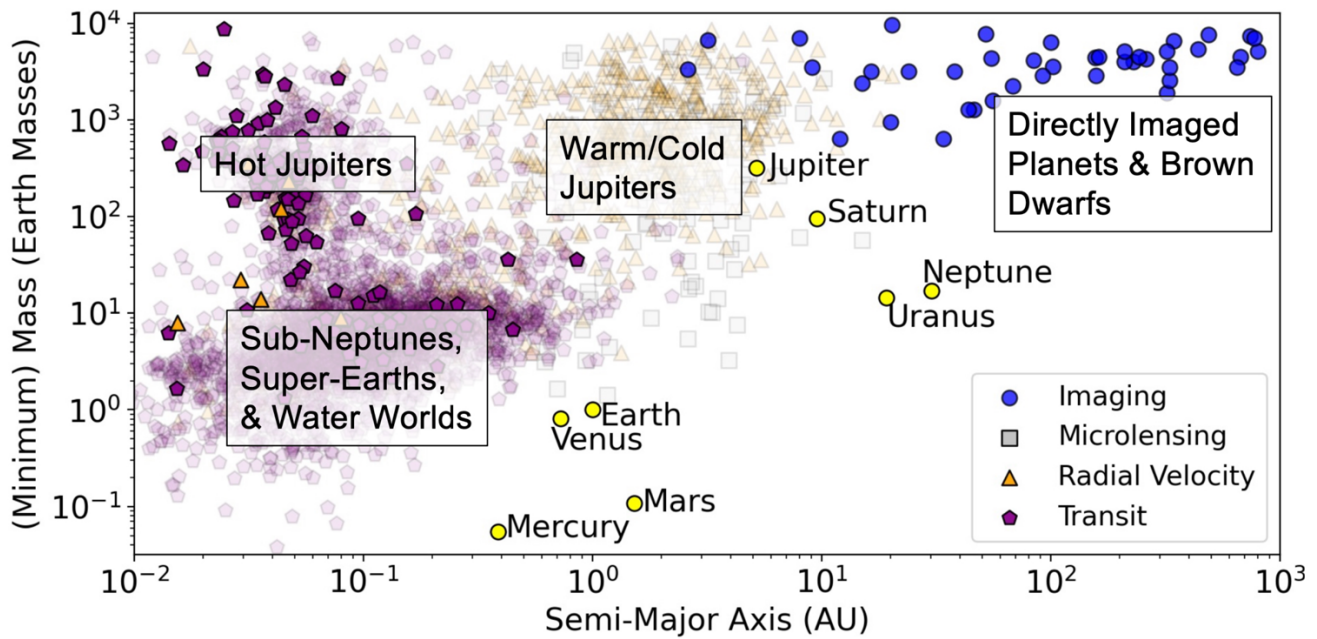

**Figure 1.** Distribution of confirmed exoplanets in mass-period space. Planets with spectroscopic measurements that constrain their atmospheric properties are shown as dark points, those without are shown as light points. This review article focuses on atmospheric characterization of *transiting* exoplanets, i.e. the dark purple pentagon symbols. Solar system planets are shown as yellow circles for context. Both the radial velocity and transit techniques are most sensitive to detecting massive planets on close-in orbits, while the direct imaging technique is most sensitive to young, self-luminous planets on relatively wide orbits. Figure adapted from Currie et al. (2023).

Current radial velocity surveys of bright nearby stars have baselines as long as $\sim 30$ years; this means that our knowledge of the occurrence rates of gas giant planets in these data sets is limited to planets with orbital semi-major axes comparable to or less than that of Saturn in our own solar system.

Exoplanets smaller than Neptune (typically defined as $< 4$ R$_\oplus$ or $\lesssim 10$ M$_\oplus$) are often found on close-in orbits around Sun-like stars. Such planets have an overall much higher occurrence than the gas giant planets: $\sim 50\%$ for orbital periods of less than 100 days (just outside the orbit of Mercury in the solar system; e.g., Fulton & Petigura 2018; Hsu et al. 2019). This population is observed to have a bimodal radius distribution, with peaks at 1.3 and 2.4 R$_\oplus$ (e.g., Fulton et al. 2017; Van Eylen et al. 2018; Fulton & Petigura 2018; Hardegree-Ullman et al. 2020; Petigura et al. 2022). The smaller planets (radii between 1.0–1.7 R$_\oplus$) have bulk densities consistent with Earth-like compositions (e.g. Lozovsky et al. 2018; Dai et al. 2019), and are therefore termed as 'super-Earths'. The larger planets (radii between 1.7–3.5 R$_\oplus$) have lower bulk densities, consistent with the presence of modest (a few percent of the total planet mass) hydrogen- and helium-rich gas envelopes (e.g., Lozovsky et al. 2018; Lee 2019; Neil et al. 2022). These planets are therefore termed as 'sub-Neptunes', although some may also have water-rich envelopes (see discussion below). The location of the bimodal radius 'gap' moves towards smaller radii at larger orbital separations (Fulton et al. 2017; Van Eylen et al. 2018; Fulton & Petigura 2018; Hardegree-Ullman et al. 2020; Petigura et al. 2022). This suggests that the gap was carved out by either photoevaporative (e.g., Owen & Wu 2017) or core-powered (e.g., Ginzburg et al. 2018; Gupta & Schlichting 2019) mass loss[6], although Lee et al. (2022) proposed that the division between the two populations might instead be largely primordial.

The order-of-magnitude occurrence rates stated above apply to planets orbiting Sun-like stars (meaning F/G/K main-sequence stars, for astronomers). These values change with decreasing stellar mass; gas giant planets are a factor of 2-3 less common around low-mass stars (e.g., Montet et al. 2014;

---

[6] In photoevaporative mass loss models the atmospheric outflow is driven by heating from high-energy (extreme ultraviolet and X-ray) stellar irradiation. In core-powered mass loss models, the heat source driving the outflow is cooling of the planetary core. However, the predicted mass loss rates in core-powered mass loss models still depend on the total irradiation received by the planet, which determines the temperature of the atmosphere and the corresponding sound speed.

Bryant et al. 2023), while small planets on close-in orbits appear to be a factor of a few more common (e.g., Dressing & Charbonneau 2015; Mulders et al. 2015; Hardegree-Ullman et al. 2019; Hsu et al. 2020).

### 1.2.3. *Composition Constraints from Bulk Density Measurements*

For planets with measured masses and radii, we can obtain a constraint on their bulk densities and corresponding bulk compositions. The bulk densities of gas giant exoplanets are relatively low, indicating that they possess thick, hydrogen-dominated gas envelopes (e.g. Thorngren et al. 2016). These bulk densities can be used to place an upper limit on the abundance of hydrogen and helium relative to heavier elements in the planet's atmosphere (often referred to as the planet's 'atmospheric metallicity' by astronomers; Thorngren et al. 2019). Exoplanets smaller than Neptune exhibit widely varying bulk densities, which reflect their varying bulk compositions. Rocky super-Earths have relatively high bulk densities, while sub-Neptunes with puffy hydrogen-rich atmospheres have much lower bulk densities (e.g. Lozovsky et al. 2018; Neil et al. 2022).

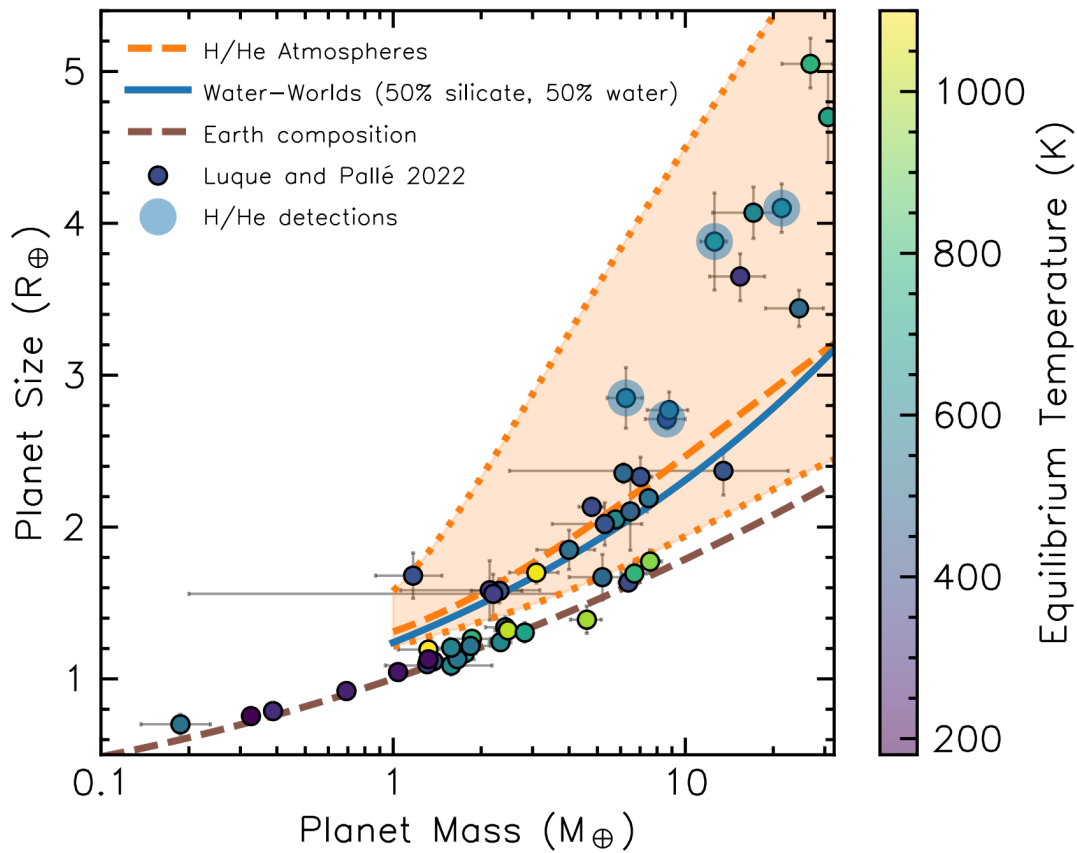

**Figure 2.** Measured masses and radii for small planets orbiting small (M dwarf) stars from Luque & Pallé (2022). This sample only includes the subset of planets with small fractional uncertainties in mass and radius. Theoretical mass-radius relations for planets with an Earth-like bulk composition (dashed brown line), half water and half Earth-like (solid blue line), and Earth-like with a few percent hydrogen-rich atmosphere (dashed orange line) are overplotted for comparison. The light orange shading denotes the range of possible hydrogen-rich atmospheres that can be retained by these planets under a range of possible starting assumptions. *Figure adapted from Rogers et al. (2023).*

Small planets with intermediate densities are more ambiguous, as their masses and radii can be equally well fit with either water-rich or hydrogen-rich envelopes (e.g., Mousis et al. 2020; Turbet et al. 2020; Aguichine et al. 2021, see Fig. 2). For lower-mass stars, which are less luminous, the water ice line is located much closer to the star. This means that even relatively close-in planets forming around low-mass stars may still be able to accrete significant quantities of ice-rich solids (e.g. Kimura & Ikoma 2022). Although some of these 'water worlds' may subsequently accrete hydrogenrich envelopes, such envelopes are more difficult to retain when they orbit low-mass stars. These stars are more magnetically active than their solar counterparts, which means that they emit more high energy photons, and also have more frequent flares and coronal mass ejections, all of which can drive atmospheric outflows (e.g., Harbach et al. 2021; Atri & Mogan 2021). It is therefore thought that planets with water-dominated envelopes may be more common around low-mass stars. Luque & Pall´e (2022) plotted all of the currently known planets orbiting low-mass stars with precisely measured masses and radii in mass-radius space and identified a sub-population of low-density planets whose densities appear to be well-matched by water-rich compositions (for alternative hydrogen-rich models, see Rogers et al. 2023). Upcoming observations of candidate water worlds using JWST will soon provide the first direct constraints on their atmospheric water content (see Section 2). These atmospheric characterization studies should provide a much clearer picture of the relative frequency of water worlds around low-mass stars.

### 1.3. *Observational Techniques for Atmospheric Characterization*

There are multiple complementary techniques that can be used to detect and characterize the atmospheric compositions of transiting extrasolar planets, as detailed below. All of these techniques leverage knowledge of the transiting planet's orbit to disentangle the combined (unresolved) light from the planet and its much brighter host star. This is distinct from the approach used to characterize directly imaged planets and brown dwarfs, whose thermal emission can be spatially resolved from that of their host stars (for a recent review, see Currie et al. 2023).

When a transiting planet passes in front of its host star, it will block more of the star's light and therefore appear larger in wavelengths at which the planet's atmosphere is strongly absorbing. Conversely, the planet will appear smaller and block less of the star's light in wavelengths at which its atmosphere is relatively transparent. This wavelength-dependent transit depth is called a 'transmission spectrum', and is the most widely used method for characterizing the atmospheric compositions of transiting extrasolar planets. We can calculate the relative size of this wavelength-dependent change in transit depth $\delta D_{tr}$ using the following expression:

$$\delta D_{tr} \simeq \frac{2R_P \delta R_P}{R_*^2}$$

(1)

where $R_P$ is the planet radius, $R_*$ is the stellar radius, and $\delta R_P$ is the wavelength-dependent change in the planet radius. The approximation holds so long as $\delta R_P \ll R_P$, which is true even for hot giants with very low densities. This change can be approximated as a multiple of the atmospheric scale height $H$:

$$\delta R_P \simeq sH = s\left(\frac{kT_{eq}}{\mu g}\right)$$

(2)

where $k$ is the Boltzmann constant, $T_{eq}$ is the predicted atmospheric equilibrium temperature, $\mu$ is the atmospheric mean molecular weight, and $g$ is the planet's surface gravity. The scaling factor $s$ typically ranges between $1-5$, with lower values more representative of weak absorption and/or atmospheres with significant aerosol opacity, and higher values more representative of strong absorption in a cloud-free atmosphere (e.g., Seager & Sasselov 2000; Miller-Ricci et al. 2009; Benneke & Seager 2012, 2013). For atmospheres with very high aerosol opacity, this signature may be completely obscured (see §3). If we assume that the star and planet both radiate as blackbodies, we can calculate the planet's predicted equilibrium temperature as:

$$T_{eq} = T_* \sqrt{\frac{R_*}{a}} \left[\frac{1}{4}(1 - A_B)\right]^{1/4}$$

(3)

where $T_*$ is the effective temperature of the host star, $a$ is the planet's semi-major axis, and $A_B$ is its Bond albedo (defined as the fraction of incident radiation that is reflected back to space; Seager 2010).

This equation assumes that the planet's atmosphere efficiently redistributes heat from the day side to the night side. In the limit of no heat redistribution (i.e., instantaneous radiative equilibrium at each longitude and latitude point), the effective hemisphere-integrated dayside equilibrium temperature can be calculated by replacing the factor of $\frac{1}{4}$ with a factor of $\frac{2}{3}$ (Hansen 2008), and intermediate values are possible between these two extremes[7]. As demonstrated by these expressions, the overall strength of absorption during the transit can vary by more than an order of magnitude when comparing hydrogen-dominated (low mean molecular weight) atmospheres to those with higher mean molecular weights (e.g., water, carbon dioxide, methane). The presence of high-altitude aerosols from photochemical hazes or condensate clouds can also attenuate the amplitude of gas absorption features by scattering the stellar photons as they pass through the atmosphere. The geometry of transmission spectroscopy means that we are primarily sensitive to the properties of the planet's atmosphere near the day-night terminator. This is particularly relevant when determining cloud properties, which can vary significantly with longitude (see Section 6). Even for clear atmospheres without significant cloud opacity, the relatively long path length of starlight passing through the planet's atmosphere means that this technique is primarily sensitive to atmospheric pressures between $0.001 - 0.1$ bars for typical hot Jupiter atmospheres observed at near-infrared wavelengths (e.g., Fortney 2005; Sing et al. 2016).

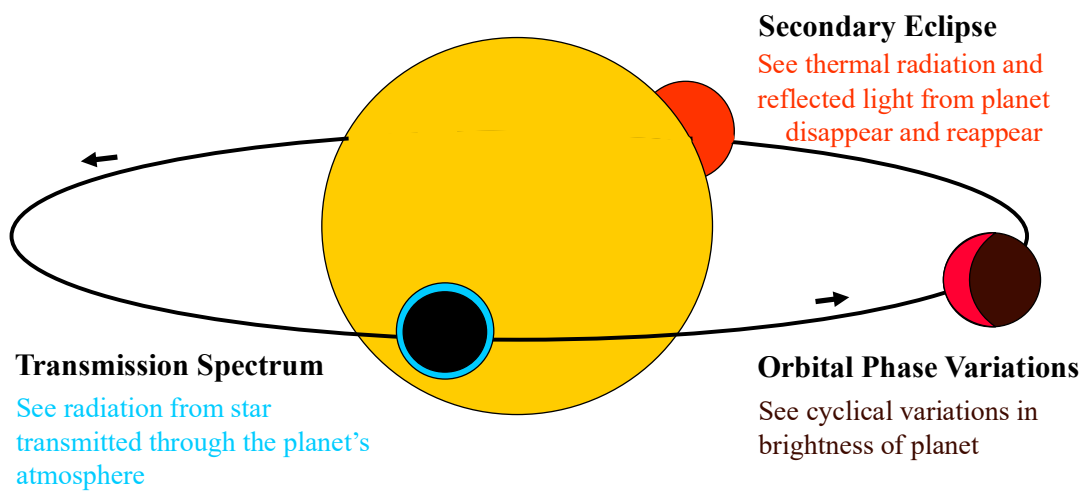

**Figure 3.** Schematic diagram illustrating three techniques that can be used to characterize the atmospheric properties of transiting exoplanets. *Adapted from a figure originally created by Sara Seager (private commun.).*

---

[7] Intermediate regimes are typically parameterized by replacing this fraction with an unknown redistribution parameter *f*. Along with the Bond albedo, this redistribution parameter can be directly constrained by infrared phase curve observations (e.g., Schwartz & Cowan 2015; Schwartz et al. 2017); see discussion below.

If we wait approximately half an orbit we can also observe the planet passing behind its host star (the 'secondary eclipse'). By measuring the relative decrease in light during this eclipse, we can determine the amount of light reflected (at optical wavelengths) or emitted (at IR wavelengths) by the planet. If we assume that the star and the planet both radiate as blackbodies and take the longwavelength (Rayleigh-Jeans) limit, we can write a simple expression for the depth of the secondary eclipse $D_{sec}$:

$$D_{sec} \simeq \left( \frac{R_P}{R_*} \right)^2 \frac{T_{eq}}{T_*}$$

(4)

The planet's emission spectrum also contains information about its atmospheric composition, as well as the average temperature as a function of pressure in its dayside atmosphere. For cloud-free hot Jupiter atmospheres observed at near-infrared wavelengths, the shorter path length of light emitted from the deeper layers of the atmosphere means that we can also potentially probe somewhat higher (a factor of a few) pressures as compared to transmission spectroscopy (e.g., Fortney 2005; Showman et al. 2009).

Close-in exoplanets are expected to be tidally locked, and as a result can exhibit large daynight temperature gradients. This means that the temperature, chemistry, and cloud properties on the daysides of these planets can differ from those measured at the terminator via transmission spectroscopy. We can obtain a global view of these atmospheres by measuring changes in the planet's brightness as a function of orbital phase (the planet's 'phase curve'). By measuring the planet's phase curve at IR wavelengths where its spectrum is dominated by thermal emission, we can map its emission spectrum as a function of longitude (Cowan & Agol 2008, 2011; Rauscher et al. 2018; Morris et al. 2022). These phase curves provide invaluable information about the atmospheric circulation patterns of tidally locked exoplanets; see Section 6 for more details. We can also spatially resolve the dayside atmosphere using a second technique called 'eclipse mapping' (Williams et al. 2006; de Wit et al. 2012; Majeau et al. 2012). This technique utilizes the measured changes in brightness during secondary eclipse ingress and egress (defined as the periods when the planet is only partially occulted by the star; for definitions of these terms see Winn 2010) to map the planet's dayside brightness as a function of longitude and latitude. This technique is complementary to phase curve observations, which can characterize the planet's night side but can only measure changes in the planet's atmospheric properties as a function of longitude.

To date, most published observations of exoplanet atmospheres have been obtained at low spectral resolution using space telescopes (Spitzer, HST, and/or JWST). Because all of these techniques rely on measurements of very small changes in the star's brightness over multi-hour timescales, it is often difficult to achieve the required stability and precision using ground-based observatories. This is because the properties of Earth's atmosphere also vary on similar timescales. However, recent advances in instrumentation on ground-based telescopes have opened up new venues for atmospheric characterization at higher spectral resolution ($R > 20,000$, where $R = \delta\lambda/\lambda$). At these resolutions, spectral features from the star, planet, and Earth's atmosphere can all be readily differentiated from one another. Crucially, the planet's spectral features are Doppler shifted by its orbital motion, while those of the star and Earth's atmosphere remain approximately constant in wavelength over several hour timescales. This means that we can use this wavelength-dependent shift to uniquely identify the absorption features from the planet's transmission or emission spectrum. Notably, this technique is not limited to transiting exoplanets and can also be used to detect spectral features in the emission spectra of non-transiting planets. For more details see the review by Birkby (2018).

### 1.4. Common Model Frameworks for Interpreting Exoplanet Spectra

When fitting transmission and emission spectra, we must necessarily make a range of simplifying assumptions in order to build simple parametric models that can be used in atmospheric retrieval frameworks. Although we are only sensitive to the atmospheric properties in a narrow range of pressures (typically $0.001 - 1$ bars for hot Jupiter atmospheres observed at infrared wavelengths at low to moderate spectral resolution), most retrievals typically assume that the inferred elemental abundances are representative of the bulk atmosphere (i.e., there is no net gradient in elemental abundances over the range of pressures, latitudes, or longitudes probed). Similarly, fits to transmission spectra often make the simplifying assumption that the atmosphere is isothermal, while fits to emission spectra typically utilize a simple parametric vertical temperature profile with up to six free parameters (e.g., Madhusudhan & Seager 2009; Line et al. 2013). Many models assume that the atmospheric chemistry is in local thermal equilibrium, or retrieve for the abundances of individual molecules assuming a single fixed abundance for each molecule as a function of pressure. For an overview of the exoplanet retrieval codes commonly in use and the corresponding assumptions made by each, see MacDonald & Batalha (2023). It is worth noting that the high quality of recent JWST observations of hot Jupiters has forced modelers to revisit many of these assumptions, some of which have proven to be too simple for the sensitivity of these new data sets. For more background on exoplanet atmosphere modeling, the reader is encouraged to refer to the following review articles: Marley & Robinson (2015), Madhusudhan (2019), and Fortney et al. (2021).

## 1.5. *High-level Scientific Questions*

Our ability to characterize transiting exoplanet atmospheres is fundamentally limited by our great distance from these systems and the fact that the planet is viewed as an unresolved object, blended with the light from its host star. Despite immense improvements in remote sensing capabilities, it is safe to say that we will never in any of our lifetimes characterize an individual exoplanet atmosphere to the degree that we have for planets within our solar system. This is a crucial piece of context for the non-astronomer to understand when formulating a realistic vision for the types of questions that exoplanet studies can address. Yet exoplanets also present an immense opportunity — that of studying a myriad of planetary systems at a *population* level. Exoplanets also provide access to types of planetary environments that do not exist in our solar system (e.g. hot Jupiters, sub-Neptunes, super-Earths, and perhaps water-worlds). A simple summary is that exoplanets allow for coarse measurements for many objects, whereas solar system studies provide detailed data on the outcome of a single instance of planet formation. Leveraging both types of information together equips us with a more complete view of planetary systems and the processes that give rise to them.

Given this context, the types of questions that exoplanet atmosphere studies aim to address are typically those that relate to bulk properties or large-scale atmospheric structure, or those that tie a collection of rough measurements to our understanding of the exoplanet population (or a subpopulation, thereof). Below, we provide an illustrative list of major open scientific questions that can be targeted through exoplanet atmospheric studies. These questions span the planet size and temperature range represented by transiting exoplanetary systems. Meaningful movement toward answering any of these questions would represent a major advance for (exo)planetary science.

- Did close-in gas giant planets form *in situ* or migrate in from farther out in the disk?

- What are the large scale atmospheric dynamics for hot Jupiters, and how do they differ with respect to solar system giant planets and young, hot, directly-imaged planets on wide orbits?

- What are the aerosols in exoplanet atmospheres made of and how do they form?

- How much hydrogen and helium gas can small planets accrete, and which planets keep (or lose) their primordial hydrogen-rich atmospheres?

- Do water worlds exist, and if so, how common are they?

- How do interactions with magma oceans shape the observed atmospheric compositions of subNeptunes and terrestrial exoplanets?

- What kinds of outgassed, high mean molecular weight atmospheres do terrestrial planets have, and what does that mean for their potential habitability?

- Which terrestrial exoplanets lose their outgassed atmospheres? What determines their total atmospheric masses?

## 2. ATMOSPHERIC COMPOSITION

### 2.1. *Composition as a Signpost of Formation and Atmospheric Chemistry*

The atmosphere is the outermost layer of a planet and the *only* component of an exoplanet that can readily have its composition directly measured using remote sensing techniques[8]. We therefore rely on observations of an exoplanet's atmosphere as a window into its history and the processes that shape its present-day state. For example, atmospheric observations can be used to inform our understanding of unseen features and processes such as surface-atmosphere interactions or interior structure. High $H_2S$ or $SO_2$ concentrations in a terrestrial habitable zone planet could be indicative of surface volcanism (Kaltenegger & Sasselov 2010); atmospheric $O_2$ and $O_3$ could signify the possible presence of surface life, especially when accompanied by disequilibrium biosignature pairs such as $CH_4$ (Lovelock 1965; Domagal-Goldman et al. 2014); and a water world might be distinguished from a sub-Neptune with a dry rocky interior via an elevated abundance of water in its atmosphere (e.g. Rogers & Seager 2010).

As with solar system planets, the present-day state of an exoplanet atmosphere is the outcome of its entire history of planet formation and evolution. By measuring an exoplanet's atmospheric composition, one can attempt to decode the processes that gave rise to that planet in the first place. On a single-planet basis, such an analysis is nearly impossible due to vast degeneracies in the range of histories that can all produce similar outcomes, in addition to fundamental uncertainties in the planet formation process and the evolution of protoplentary disks. But on a population level, we can hope to link trends in atmospheric properties to simple theories for how planets form and evolve, and anchor those theories with measurements, analogously to how our understanding of the history of our solar system stems from observations of the many bodies orbiting the Sun. Several examples for how trends in exoplanet atmospheric observations might be tied back to planet formation and evolution theories are listed, below:

---

[8] Spectroscopic characterization of rocky exoplanet surfaces might also be possible with JWST under ideal conditions (Hu et al. 2012; Whittaker et al. 2022).

- *Giant Planet Mass-Metallicity Relation:* Solar system giant planets exhibit a tight anticorrelation between their mass and atmospheric metallicity[9](Figure 4, left panel). A similar relation is predicted to be a general outcome of planet formation via core accretion, although there may also be considerable intrinsic scatter in the trend due to the stochastic nature of planet formation (Fortney et al. 2013; Venturini et al. 2016).

- *Carbon-to-Oxygen Ratios:* The composition of a planet depends on its formation location relative to various snow lines in the protoplanetary disk. The abundant volatiles oxygen and carbon are expected to be especially critical to forming planets due to their roles in delivering icy materials. Measuring the C/O ratio in exoplanetary atmospheres is therefore useful for linking present-day envelope composition to the planet's birth location and the relative import of accreting solids vs. gas during envelope formation (Oberg et al.¨ 2011; Madhusudhan et al. 2014a). It has been difficult to measure C/O in solar system giant planets because they are all cold enough for oxygen to be sequestered out of the observable atmosphere via condensation processes (e.g. Helled & Lunine 2014). Transiting exoplanets, which are typically highly irradiated, provide an excellent opportunity to directly measure atmospheric C/O without relying on model extrapolations (Madhusudhan 2012).

- *Other Elemental Abundance Ratios:* As with C/O, measuring elemental abundance ratios of various volatile and/or refractory species (e.g. Si/O, Si/C, Fe/O, etc.) provides a tracer for formation location and conditions (Piso et al. 2016; Lothringer et al. 2021; Crossfield 2023; Chachan et al. 2023). Relative abundance measurements can also be used to constrain physical and chemical processes such as condensation or transport (e.g. Gibson et al. 2022; Pelletier et al. 2023).

- *Atmospheric Composition Straddling the Sub-Neptune to Super-Earth Radius 'Gap':* A strong dip in exoplanetary occurrence for planets with radii ≈1.6 $R_\oplus$ has been explained as being the dividing line between two populations of low-mass exoplanets: rocky super-Earths and gas-rich sub-Neptunes (Fulton et al. 2017, and see discussion in Section 1.2). Theories of photoevaporative (Owen & Wu 2017) and core-powered (Ginzburg et al. 2018; Gupta & Schlichting 2019) mass loss both posit that the sub-Neptunes are planets that have succeeded in retaining their primordial nebular gas atmospheres, while super-Earths have lost their hydrogen entirely and have secondary high mean molecular weight atmospheres.

---

[9] Metallicity here and throughout this review article is defined as $(N_X/N_H)_{planet}/(N_X/N_H)_{Sun}$, where $N_X/N_H$ is the ratio of the number of some metal species (X) relative to hydrogen. Species X is selected differently across the literature, depending on what is most readily observable, leading to an inherent inconsistency in how metallicity is measured in different studies.

- *The Presence or Absence of Atmospheres on Terrestrial Exoplanets:* In the solar system, a "cosmic shoreline" in escape velocity and insolation separates bodies with atmospheres from those without (Zahnle & Catling 2017, Figure 4, right panel). Identifying whether a similar dividing line exists for terrestrial exoplanets will help to constrain the processes by which (exo)planets retain or lose their atmospheres.

In all of these cases, the trends being sought out are first-order to begin with, and the theories being tested are often highly simplified. As statistical trends in exoplanet atmospheres data are uncovered and as the data warrant it, it is only natural that these simpler ideas will give way to more complex ones, and progress will be made toward understanding the universality of the processes that shape planetary atmospheres throughout their lifetimes.

An even more direct way to constrain planet formation and evolution via exoplanet studies would be to observe exoplanets of different ages. In fact, in recent years a considerable number of exoplanets orbiting young stars (i.e. with ages ≲ 100 Myr) have been discovered (e.g. David et al. 2016; Benatti et al. 2019; David et al. 2019; Plavchan et al. 2020). Atmospheric observations of younger planets could reveal atmospheric escape or degassing processes while they are still ongoing (Zhang et al. 2022b) and might even show us what true primordial atmospheres look like. Unfortunately, the practical challenges to characterizing atmospheres of young planets are considerable. Young stars tend to be quite active. The resulting stellar variability hinders our ability to detect the minute atmospheric signatures of exoplanets orbiting these stars (Cauley et al. 2018; Hirano et al. 2020; Palle et al. 2020a; Rackham et al. 2023). We are also fundamentally limited by the number of nearby young stars that are bright enough to present sufficient SNR for atmospheric characterization studies.

Finally, measurements of atmospheric composition provide a direct indication of the chemical processes unfolding in a planet's atmosphere. For example, the measured abundances of molecules, atoms, and ions can be cross-checked against the predictions of thermochemical equilibrium for a given elemental mixture (e.g. Burrows 2014; Lodders & Fegley 2002; Schaefer & Fegley 2010). Departures from equilibrium are then attributed to disequilibrium processes such as vertical or horizontal mixing, or photochemistry (Tsai et al. 2023). Furthermore, the detection of any aerosol species (see Section 3) can be related back to the chemical and physical conditions that gave rise to them in the first place. We therefore turn to spectroscopic measurements of exoplanet atmospheric composition as a powerful tool for probing the physics, chemistry, and history of exoplanetary environments.

## 2.2. *Water, Water Everywhere*

The first molecule to be reliably detected in a large number of exoplanet atmospheres was $H_2O$. Water has many vibration-rotation absorption bands across the near-to-mid IR, and oxygen and hydrogen are cosmically abundant, making this an ideal molecule to search for. Furthermore water is stable in gas phase from ~370 – 2200 K — at lower temperatures it condenses into clouds, droplets, or ice; and at higher temperatures it thermally dissociates. Fortunately, most transiting exoplanets have temperatures within the range in which gas-phase $H_2O$ is the expectation.

The search for $H_2O$ in transiting exoplanet atmospheres from space was enabled by the installation of the Wide Field Camera 3 (WFC3) instrument on board HST during its 2009 servicing mission. WFC3 carries a grism centered on the strong 1.4 $\mu$m water absorption band with sufficient spectral resolution to resolve the shape of the band, and a novel spatial scanning procedure was developed to spread the exoplanetary spectrum across many detector pixels so as to minimize concerns about detector systematics (McCullough & MacKenty 2012). The first detection of the 1.4 $\mu$m water feature in a giant planet atmosphere with the WFC3 spatial scanning mode was made by Deming et al. (2013), and many more soon followed (e.g. Wakeford et al. 2013; Kreidberg et al. 2014b; Sing et al. 2016; Fu et al. 2017; Tsiaras et al. 2018; Changeat et al. 2022, etc; Figure 5).

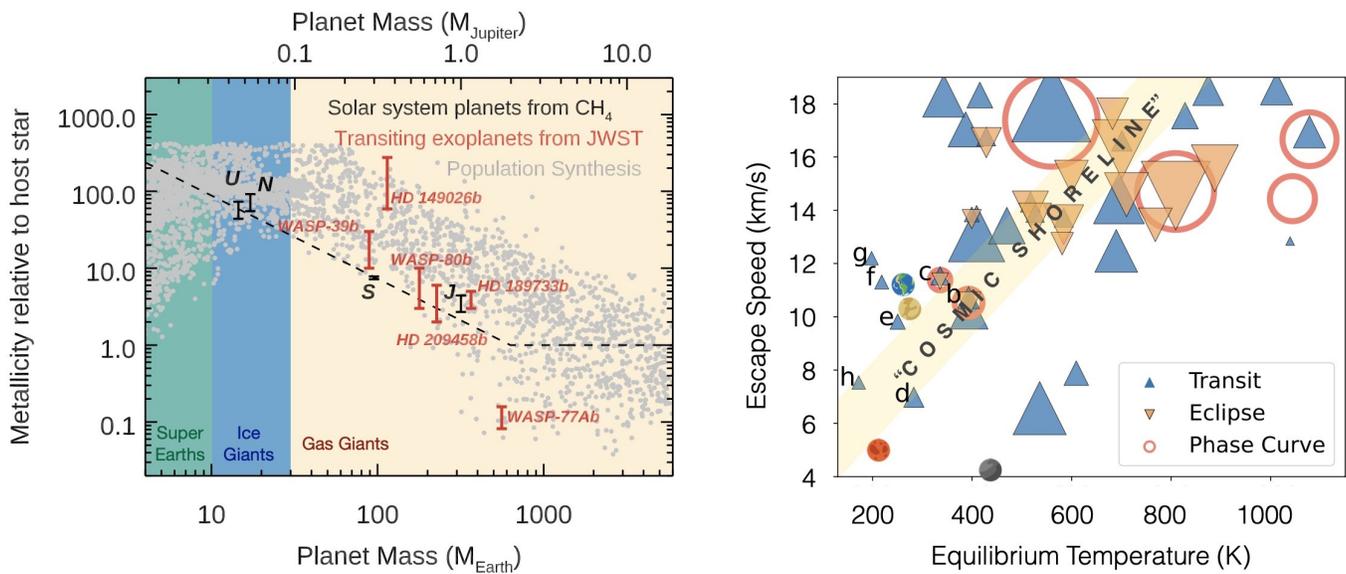

**Figure 4.** Examples of statistical comparative planetology approaches (Bean et al. 2017) to constrain planet formation and evolution processes via ensemble observations of exoplanet atmospheres. **Left:** Atmospheric metallicity vs. mass for solar system planets (black symbols) and for exoplanets that have detections of carbon- and/or oxygen-bearing species using JWST (red symbols). Overlaid are the predictions from population synthesis models from Fortney et al. (2013) showing a rise and then a plateau in metallicity as planetary mass decreases (gray dots). The solar system giant planets are observed to follow a very tight mass-metallicity correlation (dashed line), with the caveat that oxygen is undetected in the atmospheres of these planets, as it is sequestered in condensates below the photosphere. *Figure adapted from Mansfield et al. (2018).* **Right:** The "cosmic shoreline" (Zahnle & Catling 2017) is denoted (yellow diagonal band), which is an observed delineation in escape speed and insolation between solar system bodies that do and do not possess gaseous atmospheres (toward the upper left and toward the lower right of the plot, respectively). Transiting exoplanets that will be observed in Cycles 1 and 2 of JWST are over-plotted in this same parameter space. Symbol size denotes the expected S/N of a single transit or eclipse observation using the methods of Kempton et al. (2018). The letters b-h denote the planets in the TRAPPIST-1 system, which are all slated for JWST observations, and the terrestrial solar system planets are shown for reference. By identifying which terrestrial exoplanets possess atmospheres and whether they are bounded by the same "shoreline" as for the solar system, astronomers can constrain the processes by which planets lose or retain their atmospheres. *Figure courtesy of Jegug Ih.*

The ease with which the 1.4 $\mu$m water feature became detectable with HST turned this absorption band into a powerful diagnostic for the chemistry of exoplanet atmospheres. Compared to the baseline expectation of solar composition, a weaker than expected water feature in transmission can be attributed to a low water abundance, a high mean molecular weight atmosphere, or an obscuring cloud deck that mutes the underlying spectral features (Miller-Ricci et al. 2009; Benneke & Seager 2013). In thermal emission, the strength of the water feature depends on the $H_2O$ abundance and on the temperature gradient in the planet's atmosphere. Subtle differences in the shape of a spectrum resulting from each of these various scenarios can potentially be disentangled with sufficiently high S/N and broad enough wavelength coverage (Benneke & Seager 2012). By combining WFC3 measurements and longer wavelength observations with Spitzer's IRAC instrument (Figure 5), one can furthermore obtain constraints on a planet's metallicity and C/O ratio, by assuming that the atmosphere resides in a state of thermochemical equilibrium (e.g. Wakeford et al. 2018; Zhang et al. 2020). However, without simultaneous detection of the major carbon- and oxygen-bearing species in an atmosphere, these two properties remain degenerate with one another. To quantify water abundances and their associated uncertainties, novel retrieval techniques have ultimately been brought to bear on WFC3 transmission and emission spectra (e.g. Kreidberg et al. 2014b; Madhusudhan et al. 2014b; Line et al. 2016).

Ultra-hot Jupiters, which have equilibrium temperatures in excess of $\sim$2000 K, have presented a particularly interesting case for interpreting $H_2O$ detections. Weak or absent $H_2O$ features in dayside thermal emission spectra were noted for multiple ultra-hot Jupiters, and several hypothesis were posed to explain these observations (Evans et al. 2017; Sheppard et al. 2017; Kreidberg et al. 2018). Initially, retrievals were run that indicated either very low metallicities or very high C/O ratios (Sheppard et al. 2017; Pinhas et al. 2019; Gandhi et al. 2020b). The former reduces the abundances of all 'metal'-bearing species, and the latter reduces the $H_2O$ abundance by tying up nearly all atmospheric oxygen in the CO molecule. Either of these abundance patterns would be surprising though, especially since slightly cooler hot Jupiters are not observed to have similarly weak $H_2O$ features (Mansfield et al. 2021). A more natural explanation was posed by Arcangeli et al. (2018) and Parmentier et al. (2018) who pointed out that thermal dissociation of $H_2O$ at temperatures in excess of $\sim$2200 K coupled with the onset of continuum opacity from the hydrogen anion (H$^-$) around the same temperature provided a high-quality fit to available data without resorting to elemental abundance patterns that differed dramatically from the planets' host stars. Still, the precision and wavelength coverage from HST and Spitzer alone were not sufficient to unambiguously resolve the question of why the ultra-hot planets have muted water features.

On the other end of the planetary 'spectrum', sub-Neptunes and super-Earths have also been observed to have muted or absent water features in transmission. In this case, the interpretation is different because these smaller planets can have high mean molecular weight atmospheres, and they also tend to be colder planets, which makes them potentially amenable to aerosol formation.

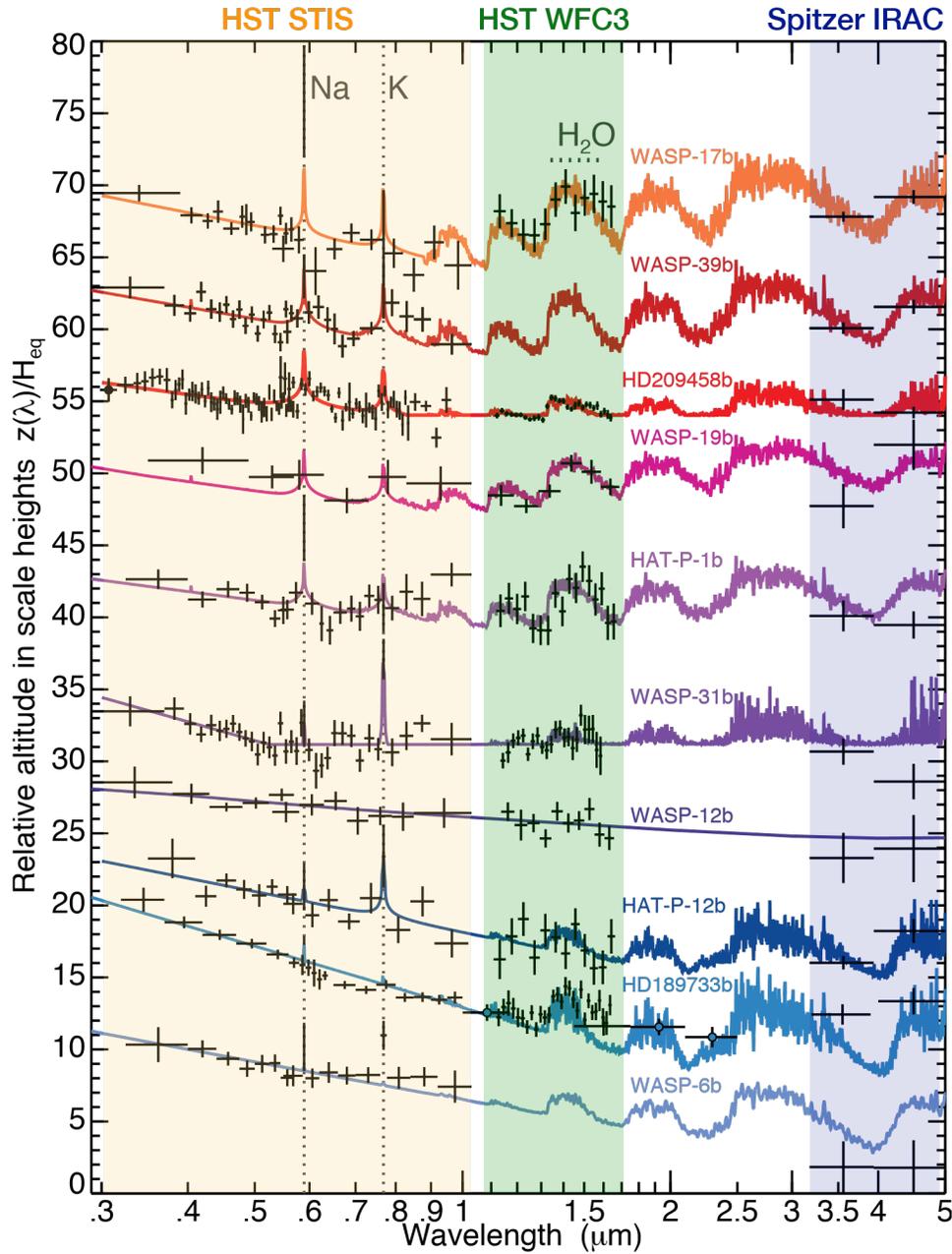

**Figure 5.** Transmission spectra for various hot Jupiters plotted in units of the planet's atmospheric scale height. Data are shown from the HST STIS and WFC3 instruments, and Spitzer IRAC channel 1 and 2 (3.6 and 4.5 $\mu m$, respectively), as indicated. The WFC3 data cover a strong water band at 1.4 $\mu m$, and STIS covers absorption lines from Na and K. The two Spitzer IRAC photometric channels are sensitive to $CH_4$, CO, and $CO_2$, although the lack of spectroscopic information over this wavelength range makes it difficult to fully constrain the atmospheric carbon chemistry. Muted spectral features and strongly sloping optical and near-IR spectra for the planets plotted toward the bottom of the figure are attributed to aerosol obscuration. *Figure adapted from Sing et al. (2016)*.

(Aerosols are not generally considered to be a major atmospheric constituent for ultra-hot planets because we do not know of any cloud or haze species that can form and persist at such high temperatures.) The spectra of rocky super-Earths will be discussed in more detail in Section 2.5. A key science question that astronomers aimed to address with the initial atmospheric observations of sub-Neptunes was to break degeneracies between 'mini-Neptune' and water world scenarios by measuring the atmospheric composition and ascertaining whether it was hydrogen- or water-dominated (Miller-Ricci et al. 2009; Miller-Ricci & Fortney 2010; Rogers & Seager 2010). Unfortunately, degeneracies between aerosols and high mean molecular weight made such distinctions extremely challenging with available instruments prior to the launch of JWST (e.g. Bean et al. 2010; Berta et al. 2012; Knutson et al. 2014; Guo et al. 2020; Mikal-Evans et al. 2021, 2023a). Perhaps the most famous among sub-Neptunes is the planet GJ 1214b, which was observed to have a staggeringly flat transmission spectra with 12 stacked transits with the HST+WFC3 instrument (Kreidberg et al. 2014a). The spectrum is so featureless that the only viable interpretation is a very thick and high-altitude layer of clouds or haze (see Section 3), which obscures any direct indications of the atmosphere below.

One challenge to interpreting any claimed detections of water in low-mass exoplanets is that many such planets that are amenable to atmospheric characterization necessarily orbit low-mass Mdwarf stars. Small host stars are required to produce large transit depths and therefore sufficiently high S/N transmission spectra. But low-mass stars also have water in their *own* spectra due to their correspondingly low temperatures. What's worse, the water is not expected to be uniformly distributed throughout the stellar atmosphere and instead to preferentially lie in cooler star-spot regions. The result is that $H_2O$ features can be spuriously imprinted on transmission spectra for planets orbiting M-dwarfs that do not originate in the planetary atmosphere but actually in the star itself (e.g. Deming & Sheppard 2017; Rackham et al. 2018; Zhang et al. 2018; Lim et al. 2023). Techniques for mitigating stellar contamination in the transmission spectra of these systems is an ongoing area of research.

Detection of $H_2O$ from the ground has also been enabled via high resolution spectroscopy. The first such measurement was made by Birkby et al. (2013) using the high-resolution CRIRES spectrograph on the Very Large Telescope (VLT) to capture water absorption lines in the dayside emission spectrum of the hot Jupiter HD189733b. As the observing techniques have matured and more near-IR high resolution spectrographs have come online, the rate of ground-based water detections has accelerated (see Figure 6). One particular advantage of high-resolution water detections over space-based measurements with HST+WFC3 is that the former are often simultaneously sensitive to oxygen- and carbon-bearing molecules, enabling direct constraints on the atmospheric C/O ratio (Pelletier et al. 2021; Line et al. 2021; Brogi et al. 2023). Such measurements have indicated C/O values for various hot Jupiters ranging from near-solar (the solar value is 0.55) to super-solar values near 1. With significant scatter in the results to-date, the implications for hot Jupiter formation are murky, but the picture should solidify in the coming years with many more direct C/O measurements enabled by JWST. Ground-based measurements of $H_2O$ have been attempted for sub-Neptunes and super-Earths as well, typically at lower spectral resolution, but to-date all have resulted in nondetections (e.g. Bean et al. 2010, 2011; Ca´ceres et al. 2014; Diamond-Lowe et al. 2018, 2020a,b).

## 2.3. *Refractory Species in Hot Jupiter Atmospheres*

Hot and ultra-hot Jupiters have high enough temperatures that most refractory species are rendered in the gas phase, and some can even be ionized via thermal or non-thermal processes. This is advantageous for exoplanet studies because it means that many elements that would otherwise be sequestered deep within a colder giant planet like Jupiter or Saturn are accessible to direct detection. Measured abundance patterns can then be compared to theories of planet formation or used to identify various chemical processes, as discussed in Section 1.5. Another goal of refractory species detections in hot Jupiter atmospheres has been to identify the optical and UV absorbers that drive thermal inversions in these planets (see Section 5 for more details). TiO and VO were initially proposed as likely species to drive stratospheric inversions in hot Jupiters (Fortney et al. 2008). More recently Lothringer et al. (2018) pointed out that a whole host of atomic metals, metal hydrides, and oxides should be in the gas phase in ultra-hot planets and would serve as even stronger optical and UV opacity sources.

Motivated by goals of measuring refractory abundances and identifying key optical absorbers, a large and increasing number of chemical species have been detected in hot Jupiter atmospheres in recent years. This has mostly been made possible by ground-based high-resolution spectrographs that observe at optical wavelengths, as well as the STIS instrument and WFC3/UVIS instrument mode aboard HST. Space-based observations with STIS and WFC3/UVIS jointly provide broad near-UV to optical wavelength coverage ($\sim$ 200–1000 nm) but only at relatively low spectral resolution, which presents a challenge for uniquely identify chemical species. For example, with low-resolution optical spectra, it has often been difficult to disentangle the causes of slopes in transmission spectra, which can be attributed to some combination of aerosol scattering (see Section 3), stellar activity, or optical absorbers (e.g. Pont et al. 2008; McCullough et al. 2014; Evans et al. 2018). At near-UV wavelengths certain ultrahot planets have been observed to have sharply increased transit depths, consistent with the presence of SiO, SH, Mg, and/or Fe, which would serve to drive thermal inversions or act as condensate cloud precursors (Evans et al. 2018; Fu et al. 2021; Lothringer et al. 2022). Individual strong lines due to atomic (e.g. Na, K; Sing et al. 2016) and ionic (e.g. $Fe^+$ and $Mg^+$; Sing et al. 2019) species have been easier to uniquely identify, albeit the line profiles are typically not fully resolved by HST, resulting in degenerate interpretations of abundances vs. broadening mechanisms. Sodium and potassium in particular have been identified in a large number of hot Jupiter spectra with STIS (see Figure 5). In some planets just one of these two species is detected, whereas others produce clear detections of both. Identifying abundance patterns in Na and K vs. fundamental parameters such as equilibrium temperature has so far been elusive.

The optical opacity 'bumps' that have been observed with HST can be fully resolved via highresolution spectroscopy in order to uniquely identify the species present. To date, well over a dozen elements and 37 individual molecular, atomic, and ionic species have been identified in hot Jupiter atmospheres with high-resolution techniques, spanning a broad portion of the periodic table (e.g. Wyttenbach et al. 2015; Hoeijmakers et al. 2018; Ehrenreich et al. 2020; Tabernero et al. 2021;

Kesseli et al. 2022; Langeveld et al. 2022; Pelletier et al. 2023; Flagg et al. 2023, Figure 6). Of these species, iron and sodium have so far proven the most readily detectable in a large number of hot and ultrahot atmospheres due to their especially strong and unique optical opacity patterns.

Much of the focus of high-resolution studies initially was on the *detection* of individual species. Papers reporting detection significances have recently been giving way to those that quantify relative and/or absolute abundances via high-resolution retrieval techniques (e.g. Gibson et al. 2020; Pelletier et al. 2021; Maguire et al. 2023; Kasper et al. 2023; Gandhi et al. 2023). Such studies have revealed a range of solar and non-solar abundance patterns. For instance, in a retrieval study of six high S/N ultrahot Jupiters, Gandhi et al. (2023) found iron abundances to be well-matched to the planets' host stars. However, other refractories such as Mg, Ni, and Cr presented more variable abundance patterns; and several species such as Na, Ti, and Ca were found to be uniformly under-abundant relative to stellar, implying some sort of depletion process such as condensation or ionization. In a detailed study of the ultrahot Jupiter WASP-76b, which measured abundances of 14 individual refractory species, Pelletier et al. (2023) similarly found abundances broadly consistent with solar (and stellar), with some notable exceptions. Elements with high condensation temperatures were found to be depleted, potentially implying condensation cold-trapping on the planet's night side, whereas Ni was over-abundant, perhaps indicating that WASP-76b accreted a differentiated planetary core during its late stages of formation. Studies such as these highlight the power of systematic investigations of gas-phase refractory elements in hot Jupiters to reveal the physics, chemistry, and history of these planets' atmospheres.

## 2.4. *The JWST Landscape*

The first JWST spectrum of a transiting exoplanet was released on July 12, 2022 as part of a handful of 'early release observations' (EROs) meant to demonstrate to the public the power of the newly commissioned space telescope[10] (Pontoppidan et al. 2022). The ~1300 K hot Jupiter WASP96b was targeted with the NIRISS instrument (Figure 7). The resulting spectrum spanning 0.6 – 2.8 μm had exactly the intended effect. It revealed a full rainbow of water features along with evidence for obscuring aerosols, and beyond that it gave the astronomical community a small taste of what was to come from exoplanet studies in the JWST era.

---

[10] https://archive.stsci.edu/hlsp/jwst-ero

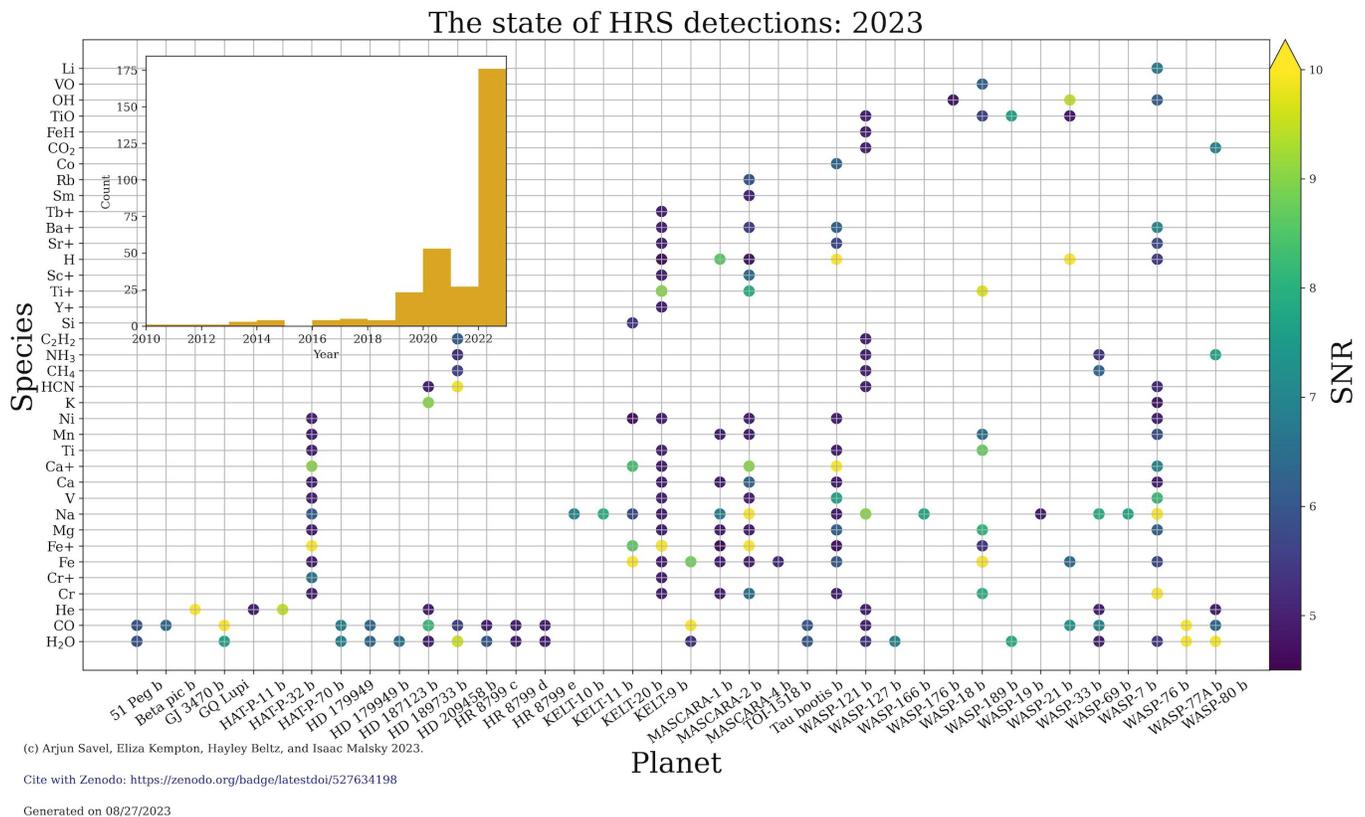

**Figure 6.** Current state of detections of ions, atoms, and molecules with high-resolution spectroscopy, as of summer 2023. The significance of each claimed detection is indicated by the symbol color. The embedded histogram shows the number of high-resolution spectroscopy atmospheric characterization papers published per year, revealing a steep acceleration. The uptick corresponds to multiple new instruments coming online as well as the maturation of the observing technique. *Figure courtesy of Arjun Savel.*

Just a month and a half later, the first peer-reviewed scientific exoplanet result from JWST revealed the striking first-time discovery of $CO_2$ in an exoplanet atmosphere (JWST Transiting Exoplanet Community Early Release Science Team et al. 2023, Figure 7). Carbon dioxide, which had previously been out of reach for spectroscopic studies due to the wavelength coverage of available instruments, was detected at a staggering significance of $26\sigma$. Chemically, the $CO_2$ molecule is especially interesting because it serves as a metallicity indicator in hot hydrogen-rich atmospheres (Fortney et al. 2010). The strong $CO_2$ absorption feature identified in the hot Jupiter WASP-39 b indicates that the planet has ~10× enhanced metallicity relative to its host star. The planet's high metallicity and low mass (relative to Jupiter), intriguingly place it right along the solar system giant planet mass-metallicity relation (Constantinou et al. 2023, Figure 4).

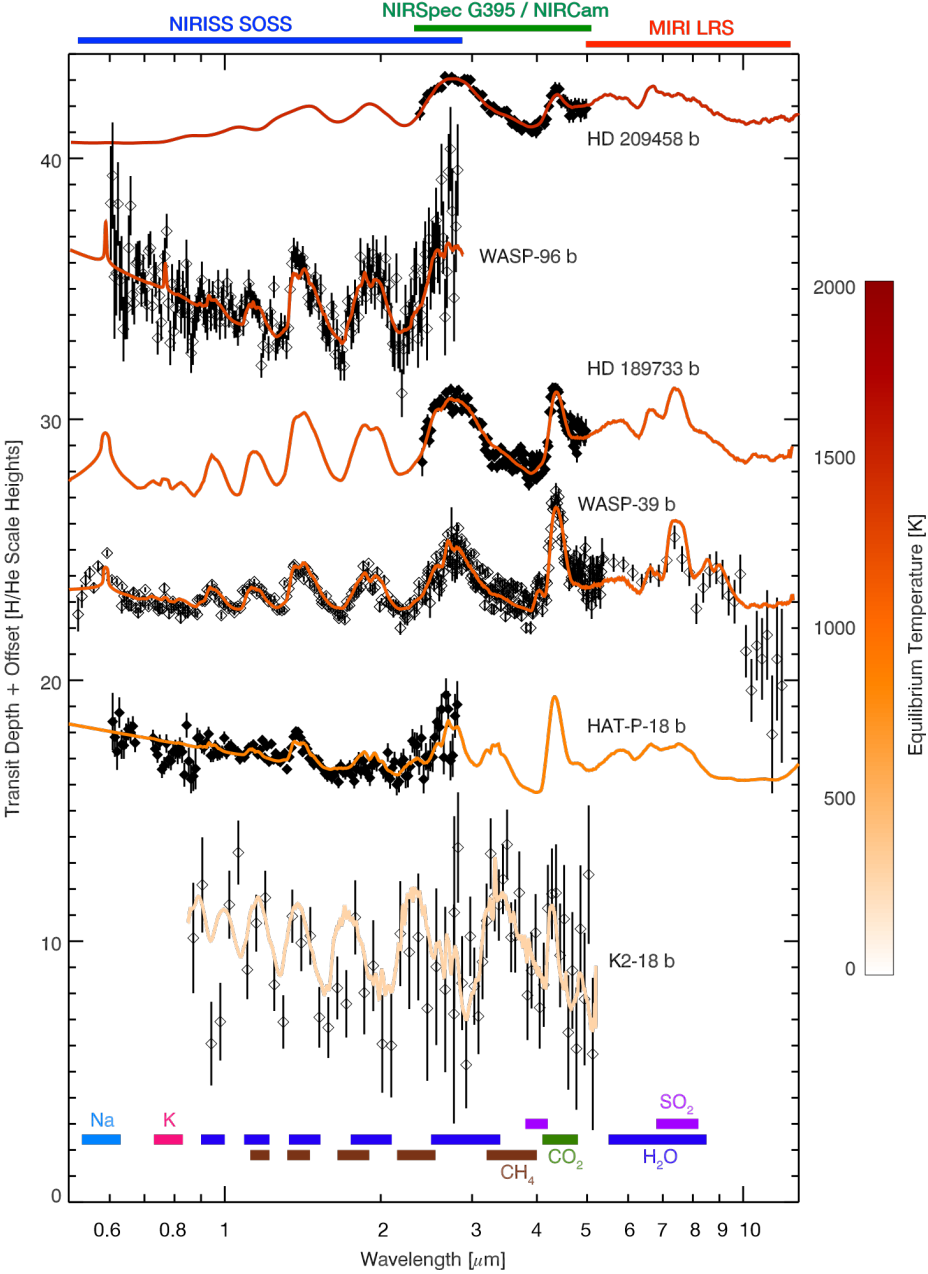

**Figure 7.** A selection of transmission spectra analyzed to-date from JWST. Wavelength coverage of the various instrument modes used for transmission spectroscopy are indicated above. Atomic and molecular opacity sources that have been identified in the planets shown are indicated below. Several other planets that have been observed with JWST but do not readily reveal any identifiable absorbers are not shown. Comparing against Figure 5, one can see the benefits of the expanded wavelength coverage and improved precision of JWST relative to HST and Spitzer for exoplanet atmospheric characterization. (*Figure data from Fu et al. (2022), Radica et al. (2023), Carter et al. (submitted), Xue et al. (submitted), and Fu et al. (in prep.).*)

Further studies of WASP-39b by the The JWST Transiting Exoplanet Community (JTEC) Early Release Science (ERS) program have since produced a full panchromatic transmission spectrum of the planet from 0.6–5.2 $\mu$m (JWST Transiting Exoplanet Community Early Release Science Team et al. 2023; Ahrer et al. 2023; Rustamkulov et al. 2023; Alderson et al. 2023; Feinstein et al. 2023, Figure 7). Spectral features from $H_2O$, $SO_2$ (Tsai et al. 2023), and CO (Grant et al. 2023; EsparzaBorges et al. 2023) have been identified, in addition to the aforementioned $CO_2$, as well as signatures of patchy aerosol coverage. The discovery of $SO_2$ at 4.05 $\mu$m is especially intriguing because this molecule was not predicted in observable amounts by any chemical equilibrium models. Instead, it is believed to be the byproduct of *photochemical* alteration of the atmosphere (Polman et al. 2023; Tsai et al. 2023). The strength of the observed feature is well-matched by hot Jupiter photochemistry models, and it is now anticipated that $SO_2$ might appear in many JWST hot Jupiter observations. The discovery of photochemically derived species opens the door to a whole host disequilibrium chemistry studies, which will be an exciting new arena for JWST.

Another possible hint of disequilibrium chemistry comes from attempts to detect methane, which is expected to be abundant in hydrogen-rich atmospheres below ~1000 K. Exoplanet atmosphere observations prior to the launch of JWST already hinted at a 'missing methane' problem, with cooler planets not showing obvious signs of methane absorption in HST or Spitzer data (e.g. Stevenson et al. 2010; Kreidberg et al. 2018; Benneke et al. 2019a), although some ground-based detections had been reported (Guilluy et al. 2019, 2022; Giacobbe et al. 2021; Carleo et al. 2022). Methane should be readily observable with JWST, as it has multiple strong absorption bands over the $1 - 8$ $\mu$m wavelength range. However, the molecule is notably absent from the transmission spectrum of the ~850 K planet HAT-P-18b with JWST's NIRISS instrument (Fu et al. 2022). Recently, methane was finally detected by JWST in yet colder planets: the ~825 K 'warm' Jupiter WASP-80b (Bell et al. 2023a) and the ~360 K sub-Neptune K2-18b (Madhusudhan et al. 2023). In the latter case, the JWST measurement resolves previous ambiguity from HST+WFC3 observations as to which gas had been detected, $H_2O$ or $CH_4$ (Benneke et al. 2019b; Tsiaras et al. 2019; B'ezard et al. 2022). The accompanying detection of $CO_2$ and non-detection of water vapor in K2-18b, also perhaps indicates the presence of a liquid water ocean below the planet's thick atmosphere (Madhusudhan et al. 2023). Further JWST observations will map out the parameter space over which methane exists in hydrogen-rich planetary atmospheres and will hopefully hint at the underlying mechanisms behind the missing methane problem such as hot planetary interiors coupled with efficient vertical mixing (Fortney et al. 2020), horizontal quenching (Cooper & Showman 2006; Zamyatina et al. 2023), or photochemistry (Line et al. 2011; Miller-Ricci Kempton et al. 2012).

It is still early days for JWST, and the observatory has just begun to reveal its prowess in characterizing exoplanet atmospheric composition. Along with metallicities, the reliable measurement of C/O ratios in exoplanet atmospheres has been highly anticipated, enabled by the broad wavelength coverage of the JWST instruments. For example, the $0.6 - 12$ $\mu$m wavelength range covered by the JWST exoplanet instrument suite spans spectral features from $H_2O$, $CH_4$, $CO_2$, and $CO$, which allows for direct measurement of the atmospheric C/O ratio under the assumption of thermochemical equilibrium (e.g. Batalha & Line 2017). The first contstraints on metallicities and C/O ratios reveal the diverse outcomes of planet formation processes (Figures 4 and 8). Derived metallicities in hot Jupiters range from sub-solar (August et al. 2023) to highly super-solar (Bean et al. 2023). Whereas WASP-39b is found to lie directly along the solar system mass-metallicity relation, several other planets do not, implying either that this is not a universal correlation for giant planets, or that there is considerable scatter in the underlying trend. Measurements of C/O ratios for hot Jupiters have have also recovered a range of values from sub-solar (JWST Transiting Exoplanet Community Early Release Science Team et al. 2023; Coulombe et al. 2023; August et al. 2023), to solar (Radica et al. 2023), to super-solar (Bean et al. 2023). Published hot Jupiter studies with JWST are still in the small number statistics regime. However, planned observations of several dozen such planets with JWST in its first two years establishing whether abundance patterns align with specific theories of giant planet formation.

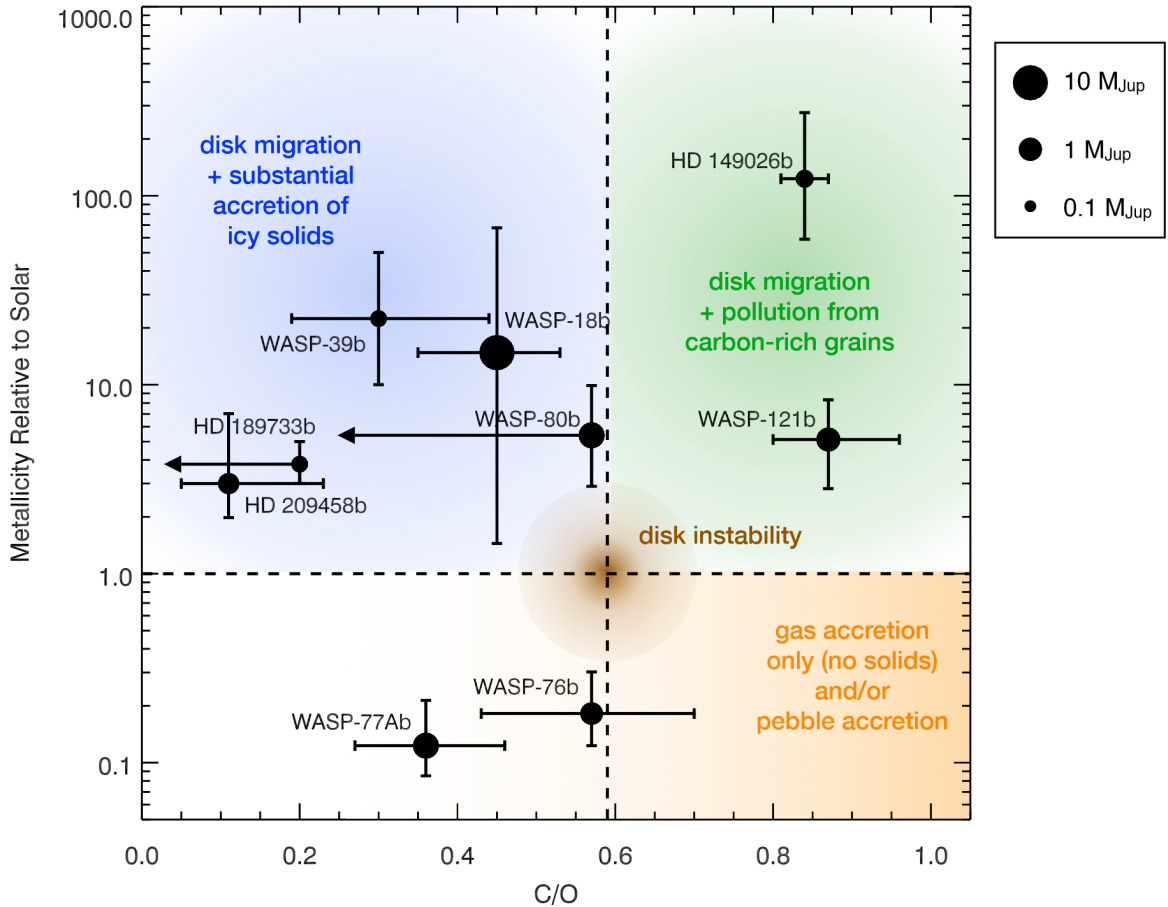

**Figure 8.** Metallicity vs. C/O ratio for all planets with measurements of both carbon- and oxygen-bearing species from JWST or high-resolution ground-based spectrographs as of the writing of this article. Formation scenarios consistent with different combinations of metallicity and C/O are a summary of work described in Öberg et al. (2011), Madhusudhan et al. (2017), Booth et al. (2017), and Reggiani et al. (2022). (*Figure data from Bean et al. (2023), Brogi et al. (2023), August et al. (2023), Bell et al. (2023a), Xue et al. (submitted.), Fu et al. (in prep.), Welbanks et al. (in prep.), Pelletier et al. (in prep.), and Mansfield et al. (in prep.).*)

Another arena in which JWST has already made its mark is to resolve previous ambiguity over the atmospheric composition of ultrahot Jupiters (see Section 2.2). Whereas with HST alone it had been challenging to determine the cause of weakened $H_2O$ features in ultrahot Jupiter emission spectra, the wavelength coverage and precision of JWST data for the planet WASP-18b has allowed for a robust measurement of the planet's underlying atmospheric composition (Coulombe et al. 2023). The NIRISS secondary eclipse spectrum of WASP-18b clearly shows evidence for weakened, but still significant, $H_2O$ features in emission. The detailed shape of the spectrum is best-fit by models with near-solar metallicity, sub-solar C/O, $H^-$ continuum opacity, and water depleted in the observable atmosphere via thermal dissociation. This composition is in line with 'vanilla' predictions of an unaltered nebular gas atmosphere in thermochemical equilibrium and rules out more exotic scenarios. Attempts to characterize even smaller exoplanets with JWST are also just beginning in earnest.

Terrestrial planets will be discussed in more detail below, but the first JWST investigations of subNeptunes are also taking shape. Following on years of ambiguous characterization of sub-Neptunes that have produced degenerate interpretations of atmospheric composition and aerosols (see Sections 2.2 and 3), the first phase curve observation of a sub-Neptune (the planet GJ 1214 b) has revealed clear evidence that the planet has a high mean molecular weight atmosphere (Kempton et al. 2023). The planet's dayside and nightside thermal emission spectra additionally show spectroscopic signs of $H_2O$ and perhaps $CH_4$ (the two are partially degenerate with one another over the mid-IR wavelength range observed). The derived composition of the planet is consistent with GJ 1214b either being a water world or 'gas dwarf', i.e. an initially hydrogen-rich planet that has experienced considerable loss of lighter elements throughout its lifetime. Approximately 20 additional sub-Neptunes are scheduled for transmission spectrum observations with JWST during Cycles 1 and 2[11], opening the door to further compositional characterization of this intriguing class of planets, so long as spectral features are not entirely obscured by aerosols.

### 2.5. *The Challenge of Terrestrial Planets*

Terrestrial planets are especially challenging targets for atmospheric characterization due to their small sizes and also the expectation that their atmospheres will typically have high mean molecular weight, which reduces the size of spectral features observed in transmission (see Section 1.3 and also a review article by Wordsworth & Kreidberg (2022) for more background on terrestrial exoplanets). The first attempts to measure the atmospheric composition of rocky exoplanets typically resulted in non-detections of spectral features, which in turn could rule out cloud-free hydrogen dominated atmospheres but left open a wide range of plausible atmospheric compositions and cloud properties (e.g. de Wit et al. 2016, 2018; Diamond-Lowe et al. 2018, 2020b, 2023; Mugnai et al. 2021; Libby-Roberts et al. 2022). To this day, there have not yet been any robust detections of atmospheric species in terrestrial exoplanets. The small number of works that have claimed the detection of atmospheric gases for rocky planets via transmission spectroscopy have been called into question or have not been reproduced (e.g. Southworth et al. 2017; Swain et al. 2021).

---

[11] JWST (and HST) observations are scheduled in annual cycles. Cycle 1 can therefore be thought of as the first year of JWST operations, Cycle 2 as the second year, etc.

An alternative approach to characterizing rocky planet atmospheres was first demonstrated by Kreidberg et al. (2019) and Crossfield et al. (2022). The former measured the phase curve of the terrestrial exoplanet LHS 3844b, and the latter observed the secondary eclipse of GJ 1252 b, both with the Spitzer Space Telescope. The goal of both observations was to determine whether the planet in question has a thick atmosphere or is an airless barren rock. The technique is discussed in more detail in Section 5.4, but briefly, the premise is that, for tidally-locked exoplanets (as is expected to be the case for these and most other terrestrial planet atmospheric characterization targets orbiting M-dwarfs), an atmosphere serves to transport heat away from the planet's hot dayside to its colder nightside (Seager & Deming 2009; Koll 2022). In both cases, the planets' high observed dayside temperatures were found to be consistent with the lack of a substantial atmosphere, although Earththickness 1-bar atmospheres could not be ruled out. The inferred limits on atmospheric thickness are also composition-dependent due to the differing abilities of various gases to absorb light and transport heat, governed by their wavelength-dependent opacities (Whittaker et al. 2022; Ih et al. 2023; Lincowski et al. 2023). With Spitzer, such dayside thermal emission measurements were only possible for the few hottest and most favorable targets, but JWST has much expanded capabilities in this arena (Koll et al. 2019).

The large aperture and IR observing capabilities of JWST have long promised to extend the parameter space of observable exoplanet atmospheres to terrestrial planets (e.g. Deming et al. 2009; Beichman et al. 2014; Batalha et al. 2015). Nearly 30 such planets (i.e. rocky super-Earths to subEarths) are already approved for observation in Cycle 1 and 2 of JWST operations. The list of planned observations includes all 7 of the planets in the TRAPPIST-1 system, as well as numerous terrestrial planets orbiting earlier (larger, warmer, and more massive) M stars, and a handful of ultrashort-period (USP) rocky planets with periods shorter than 1 day orbiting G, K, and M stars. The TRAPPIST-1 system is of particular interest for habitability studies aiming to identify biosignature gases because the late (i.e. small and cool) M-dwarf host star brings the habitable zone to very short orbital periods and produces large transit depths and thus atmospheric signal sizes (e.g. Barstow & Irwin 2016; Krissansen-Totton et al. 2018; Lustig-Yaeger et al. 2019). TRAPPIST-1 e, f, and g are all potentially habitable environments, and are largely seen as the best prospects for characterizing potentially habitable worlds within the next decade (Gillon et al. 2017).

The first year of JWST data has so far been marked by more non-detections of terrestrial atmospheres, but now with the vastly improved capabilities of the new facility, such measurements are more meaningfully constraining. Transmission spectra to-date are consistent with flat lines (LustigYaeger et al. 2023), or with the possibility that spectral features are caused by the host star and not the planet (Moran et al. 2023; Lim et al. 2023). Thermal emission studies of terrestrial planets (so far limited to TRAPPIST-1 b and c) have been more revealing. As with previous Spitzer thermal emission studies, the goal with these observations has been to measure the planets' dayside temperatures and infer the presence or lack of an atmosphere. For both planets, the measured dayside temperatures are again consistent with no atmosphere being present (Greene et al. 2023; Zieba et al. 2023). The mid-IR capabilities of JWST has allowed for these measurements to be made at much longer wavelengths than Spitzer could access. By observing at 15 microns in the center of an expected strong $CO_2$ absorption band, the JWST measurements can rule out very thin atmospheres (for TRAPPIST-1 b down to even Mars thickness), under the assumption that $CO_2$ would be a dominant gas in any moderately-irradiated terrestrial environment (Ih et al. 2023; Lincowski et al. 2023).

Still, a key promise of JWST is to deliver spectra of smaller and cooler exoplanets than what was previously possible with HST. In light of the flat-line spectra and dayside thermal emission results, a question that has supplanted the characterization of rocky exoplanets has been to identify whether such planets possess atmospheres at all. Figure 4 (right panel) shows that Cycle 1 and 2 JWST targets cover the parameter space of planets that would and would not be expected to host atmospheres, based on solar system considerations. If some of the less-irradiated and/or higher surface gravity terrestrial exoplanets are found to have atmospheres, multiple modeling studies have shown JWST's capabilities to spectroscopically characterize such environments under optimal conditions of minimal cloud obscuration, large scale heights, and stacking multiple transits to improve S/N (e.g. Barstow & Irwin 2016; Morley et al. 2017; Batalha et al. 2018; Krissansen-Totton et al. 2018; Lustig-Yaeger et al. 2019; Fauchez et al. 2019; Suissa et al. 2020; Pidhorodetska et al. 2020). For the subset of rocky planets without atmospheres, mid-IR emission spectroscopy measurements with JWST offer an exciting opportunity to characterize their surface compositions for the first time (e.g. Hu et al. 2012; Whittaker et al. 2022; Ih et al. 2023).

## 3. AEROSOLS

### 3.1. *Terminology and Background*

Aerosols in this work are defined to be any kind of particle suspended in a gaseous atmosphere, regardless of their composition or formation pathway [12]. Aerosols can be broken up into sub-categories: clouds are defined as solid particles or liquid droplets formed by *condensation* processes, hazes are involatile particles produced by chemical (and often photochemical) processes, and dust is made up of solid particles suspended in an atmosphere that originated elsewhere (e.g. particles kicked up from the surface or those that originated from a meteorite breaking up as it entered a planet's atmosphere). These are all process-based definitions. If the formation mechanism for the particles in question is unknown, we revert to the blanket term 'aerosol'. We warn the reader that some published papers in the exoplanetary literature employ the term 'haze' when referring to small particles ($\lesssim 1$ $\mu$m) and 'clouds' when referring to larger particles, but we prefer the process-based definitions for the physical insight they bring.

All solar system planets and moons with significant atmospheres have some sort of aerosol layer. For example, Earth has water clouds, surface dust, and technology-derived haze (i.e. smog). Venus has sulfuric acid clouds and haze. Titan has clouds and haze formed from organic compounds. It stands to reason that exoplanets too should have aerosol layers and that these will be a fundamental component of their atmospheres, governing energy balance, thermal structures, and observed spectra; as is the case for the solar system planets. As we will see, there is indeed plentiful observational evidence for exoplanet aerosols. For a more detailed review of aerosols in exoplanet atmospheres, we refer the reader to a recent article by Gao et al. (2021).

---

[12] This definition and those that follow for clouds, haze, and dust are all attributed to an online article written for the Planetary Society by Sarah Hörst: https://www.planetary.org/articles/0324-clouds-and-haze-and-dust-oh-my.

The observational signatures of clouds and hazes in exoplanet transmission spectra are primarily muted (or absent) spectral features or strong blue-ward spectral slopes at optical wavelengths[13]. These come about due to the propensity of the aerosol particles to scatter or absorb starlight. Rayleigh-like scattering slopes arise for small particles, whereas flatter spectra result when particle sizes are larger or the clouds are very thick. The aerosol species themselves also have their own spectral signatures, but these are typically weak features with wavelength-dependent shapes that depend on particle size distribution (e.g. Wakeford & Sing 2015). This results in degeneracies among spectra associated with distinct aerosol species, making the aerosol composition difficult to uniquely constrain spectroscopically. In thermal emission, the signatures of aeorols tend to be even more subtle. Aerosol layers can impact the thermal structure of an atmosphere (for example, causing thermal inversions in the case of very absorptive clouds or hazes, thus altering the shape of the planet's emission spectrum; Arney et al. 2016; Morley et al. 2015; Lavvas & Arfaux 2021), and optically thick aerosols can mute spectral features; but these effects are not uniquely attributable to clouds and therefore can be challenging to interpret. The result is that one can often tell from an observation that aerosols are present, and inferences can be made about the vertical distribution of the clouds or haze, but concluding anything robustly about the aerosol composition on a planet-by-planet basis is exceedingly difficult. Forward models are useful to motivate which types of aerosols are consistent with a specific observation, providing probabilistic arguments on the aerosol composition. This approach can be especially powerful at the population level.

---

[13] Additionally, since many cloud species contain oxygen, the rainout process can alter the C/O of the gas-phase atmosphere, which can impact abundance interpretations if not properly accounted for (e.g. Helling et al. 2019).

Oftentimes in the exoplanet literature, aerosols are treated as a 'nuisance' parameter, due to their impact of hindering the detection the underlying gaseous atmosphere. Modeling the complex microphysics and chemical processes that lead to aerosol formation is a challenging task, so parameterized studies that reduce the aerosols to as few defining properties as possible are common (e.g. Ackerman & Marley 2001; Benneke & Seager 2012). Yet it is only by understanding the aeorols themselves, including their composition, formation, and optical properties, that we can gain a holistic picture of the planetary atmosphere in question. Studies of exoplanet aerosols additionally provide us with unique laboratories for probing cloud and haze formation in conditions that are not accessible within the solar system.

### 3.2. *Hot Jupiter Aerosols*

For hydrogen-rich atmospheres hotter than $\sim$1000 K, the types of clouds that are able to form due to condensation processes are those that are more commonly thought of as refractory species. For example, based on chemical equilibrium calculations one would expect clouds of Fe, Ni, $Al_2O_3$, $Mg_2SiO_4$, $TiO_2$, and MnS for a solar-composition gas mixture (e.g. Burrows & Sharp 1999; Mbarek & Kempton 2016; Woitke et al. 2018; Kitzmann et al. 2023). Because such clouds are expected to only form at very high temperatures, and they incorporate trace species, it can be tempting to ignore the impacts of aerosols on hot Jupiter studies. Yet it was shown early on that transmission spectroscopy geometry, specifically the oblique geometric path taken by stellar photons through the exoplanetary atmosphere on their way to the observer, can result in cloud optical depths considerably in excess of unity, even for trace species (Fortney 2005). Obscuration by clouds was one explanation immediately put forth for the weaker than expected sodium absorption signal seen in the very first exoplanet transmission observation (Charbonneau et al. 2002). These early studies indicated that cloud modeling would need to be an integral component of interpreting hot Jupiter atmospheric observations.

The presence of aerosol layers has been inferred from multiple hot Jupiter transmission spectroscopy studies, starting with the benchmark planet HD 189733b (Pont et al. 2008). For that planet, a strong spectral slope over optical wavelengths accompanied by non-detections of sodium and potassium, which should have been present under clear atmosphere conditions, constituted strong evidence for cloud or haze obscuration. However, later work showed that the optical slope could equivalently be the signature of unocculted starspots on the surface of the planet's active host star, leading to an ambiguity in how to interpret the observational result. Since then, many other hot Jupiters have revealed optical spectral slopes and/or muted spectral features over IR wavelengths, indicating that aeorosol coverage is a likely culprit across the population (e.g. Sing et al. 2016; Wakeford et al. 2017, and see Figure 5). Other observational indications of clouds in hot Jupiters come from optical and IR phase curve observations, which will be discussed further in Section 6.2.

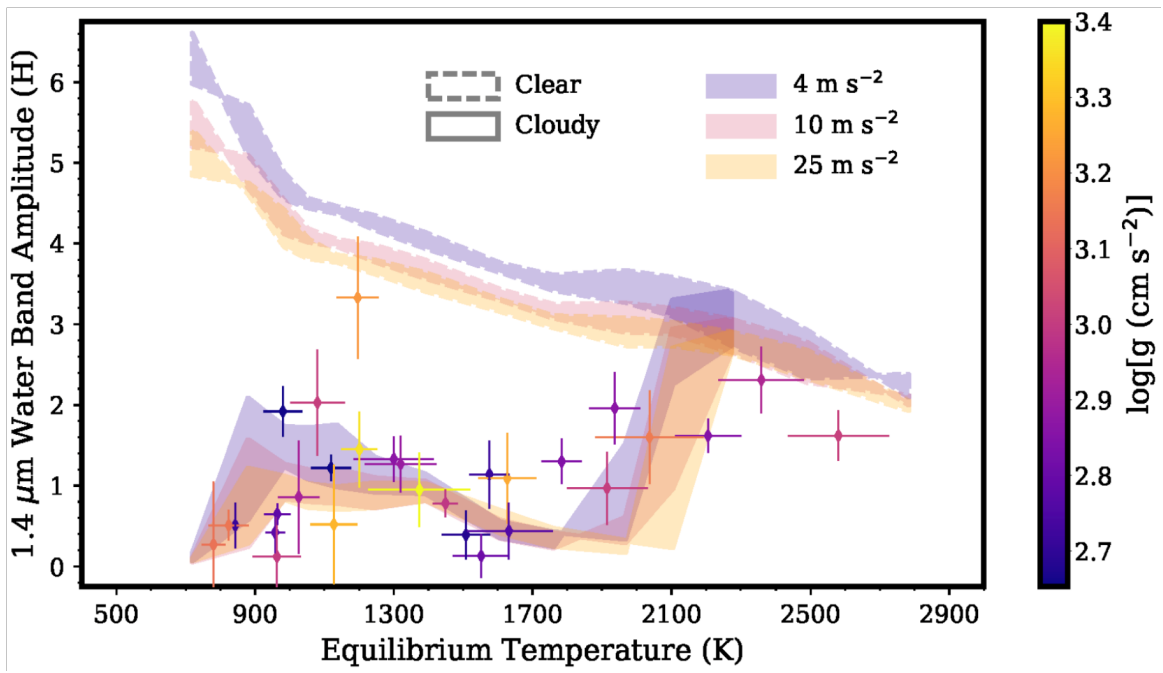

**Figure 9.** Clear and cloudy atmosphere model tracks compared with transmission spectroscopy measurements of the 1.4 $\mu$m water feature amplitude as a function of planetary equilibrium temperature for transiting gas giant planets. Hot Jupiter transmission spectra for planets colder than ~2100 K generally have weaker absorption features than what is predicted for clear solar-composition gas mixtures. The cloud microphysics models shown here provide a good overall fit to the observed trend. In addition to cloud condensation, the models include the formation of hydrocarbon hazes, which increasingly dominate at equilibrium temperatures below 950 K. The clouds are primarily formed from $Mg_2SiO_4$, with minor contributions from $Al_2O_3$ and $TiO_2$. Some variation in the degree of aerosol coverage is expected based on surface gravity, metallicity, and C/O ratio, which is likely driving the intrinsic scatter in the observed data points. *Figure from Gao et al. (2020).*

On a population level, the aerosol composition and formation mechanism can become more apparent. By comparing the strength of transmission spectral features to a suite of aerosol microphysics forward models, Gao et al. (2020) argued that the muted spectral features for transiting gas giant planets are primarily caused by silicate clouds for planets hotter than 950 K and hydrocarbon hazes for cooler planets (Figure 9). Their argument hinges on the relatively high abundances of Si, Mg, and C, their three main aerosol-forming species, and that other species of comparable abundance (e.g. Fe and Na) don't readily form clouds due to high nucleation energy barriers inhibiting particle formation.

JWST is expected to improve our understanding of hot Jupiter aerosols by providing higherprecision spectra and broader wavelength coverage, allowing for degeneracies to be broken between aerosol coverage and high mean molecular weight or non-solar abundance patterns (Batalha & Line 2017). Already this enhanced precision has led to the finding of *partial* cloud coverage of the terminator of WASP-39b due to subtle departures in the shape of its transmission spectrum from a fullyclouded planet (Feinstein et al. 2023). The access to longer wavelengths with JWST also presents the opportunity to directly measure spectral features from aerosols (e.g. Wakeford & Sing 2015), such as the silicate features that have been observed in mid-IR spectra of brown dwarfs and directly-imaged giant planets (e.g. Burningham et al. 2021; Miles et al. 2023). As for high-resolution spectroscopy studies, the signatures of aerosols are difficult to distinguish because high resolution data processing techniques typically remove the spectral continuum, which is where most of the aerosol information is contained (e.g. Snellen et al. 2010; de Kok et al. 2013). An advantage of high-resolution studies though is that the sharply peaked cores of spectral lines tend to extend above cloud decks, resulting in an ability to measure gas-phase composition even for aerosol enshrouded planets (Kempton et al. 2014; Hood et al. 2020; Gandhi et al. 2020a).

Another particularly promising avenue for further constraining aerosol composition in hot Jupiter atmospheres is by using 3-D diagnostics to determine *where* on the planet (i.e. as a function of longitude and/or latitude) the aerosols are located. The aerosol spatial distribution and the physical conditions derived at those locations (e.g. temperature, UV irradiation, wind speeds) can then be directly linked to a proposed aerosol formation mechanism and composition. Such analyses can be accomplished on high S/N spectra and phase curves from JWST or high-resolution spectra from ground-based telescopes (e.g. Kempton et al. 2017; Ehrenreich et al. 2020; Espinoza & Jones 2021; Parmentier et al. 2021; Roman et al. 2021). We discuss 3-D diagnostics for aerosols further in Section 6.2.

### 3.3. *Aerosols in Sub-Neptunes and Super-Earths*

Because they tend to orbit smaller stars and thus have cooler temperatures, transiting planets smaller than Neptune are especially likely to host aerosol layers. This was heavily implied by the first investigations of sub-Neptune exoplanets, which returned featureless transmission spectra (e.g. Bean et al. 2010; Berta et al. 2012; Knutson et al. 2014). To date, most flat and muted sub-Neptune and super-Earth transmission spectra are consistent with either aerosols or high mean molecular weight atmospheres, leading to degenerate interpretation. However, in the case of the planet GJ 1214b, the data were obtained at high enough precision by stacking multiple transits with HST that the degeneracy could be broken, and a thick aerosol layer remains as the only viable explanation (Kreidberg et al. 2014a). JWST should similarly provide the precision to break the aerosol vs. mean molecular weight degeneracy in a *single* transit for many sub-Neptunes, allowing for improved aerosol characterization for smaller planets.

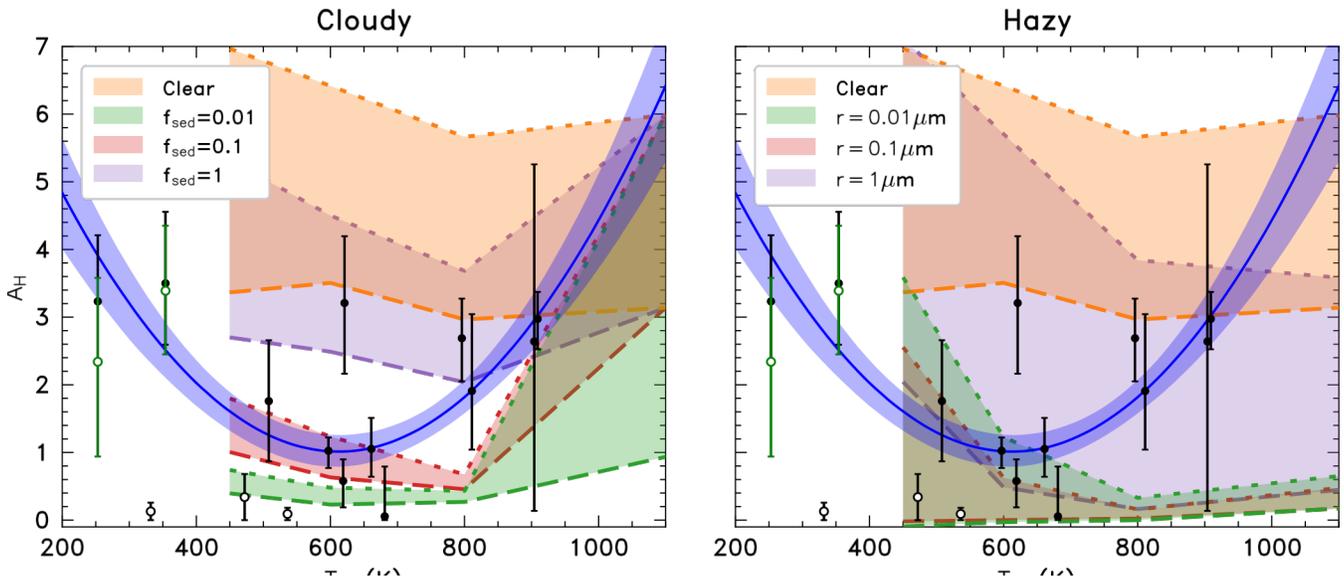

**Figure 10.** Strength of the 1.4 $\mu$m water feature in units of scale heights for transmission spectra of planets 2−6 $R_\oplus$ in size. The blue parabola is a best-fit second order polynomial trend, implying a minimum in the strength of transmission spectral features around an equilibrium temperature of 600 K, perhaps indicating the conditions for maximal aerosol coverage. **Left:** Colored lines indicate clear and cloudy models from Morley et al. (2015). The dotted and dashed contours are for 100× and 300× solar metallicity, respectively. **Right:** Colored lines indicate clear and hazy models from Morley et al. (2015). The hazy models include soot hazes only. The dotted and dashed contours are for 1% haze precursor conversion into soots while the dashed contours are 10%. *Figure courtesy of Yoni Brande.*

Even without an unambiguous detection of aerosols on a planet-by-planet basis, the ubiquity of flattened transmission spectra and indications that the flatness correlates with planetary equilibrium temperature (Crossfield & Kreidberg 2017; Libby-Roberts et al. 2020; Gao et al. 2021) point to aerosol coverage being a defining characteristic of sub-Neptune exoplanets. Hints that the aerosol coverage may clear at high temperatures ($\gtrsim$ 900 K) and lower temperatures ($\lesssim$ 400 K) provide hints as to the dominant particle composition and formation pathway (Figure 10).

As for the aerosol formation mechanism, it is hypothesized that hydrocarbon-based hazes readily form in hydrogen-rich sub-Neptune exoplanets below a temperature of ~850 K (e.g. Morley et al. 2015; Kawashima & Ikoma 2019). Under such conditions, methane is expected to be plentiful in chemical equilibrium. Methane is readily photolyzed by the UV radiation from the host star, producing a rich collection of higher-order hydrocarbons that can continue to polymerize and ultimately form large involatile haze particles (e.g. Miller-Ricci Kempton et al. 2012; Morley et al. 2013; Kawashima & Ikoma 2018; Lavvas et al. 2019). This is analagous to how we believe hydrocarbon 'tholin' haze forms in Titan's atmosphere. The propensity for hazes to form under sub-Neptune conditions is supported by lab work, in which an ensemble of gases are irradiated by a UV or plasma energy source, and the resulting solid particles are collected and analyzed (Hörst et al. 2018; He et al. 2018). Interestingly, lab studies are also able to form hazes in gas mixtures without methane (He et al. 2020), indicating that haze formation in exoplanet atmospheres may come about via diverse chemical pathways that have yet to be characterized.

Candidates for condensate clouds in sub-Neptunes include sulfides (ZnS, $Na_2S$), sulfates ($K_2SO_4$), salts (KCl), and graphite (Miller-Ricci Kempton et al. 2012; Morley et al. 2013; Mbarek & Kempton 2016). These are the expected equilibrium condensates over the temperature range of most sub-Neptunes studied to date, although some of these species may not form clouds due to their high nucleation-limited energy barriers (Gao et al. 2020). Arguments for haze being dominant over clouds in sub-Neptune atmospheres also hinge on how thick and high up the aerosols layers must be in order to fully flatten transmission spectra, especially in the case of the well-studied planet GJ 1214b. It is difficult to build models in which low-abundance species such as ZnS or KCl are able to provide sufficient opacity to match existing observational data (Morley et al. 2013, 2015).

Recent JWST phase curve observations of GJ 1214b have thrown another surprise into our evolving understanding of sub-Neptune exoplanets. The very high measured Bond albedo of the planet based on its global thermal emission ($A_B \approx 0.5$) implies that the planet's aerosol layer is highly reflective (Kempton et al. 2023). This is in tension with our understanding of hydrocarbon hazes (e.g. soots and tholins), which are primarily believed to be dark and absorptive (Khare et al. 1984; Morley et al. 2013, 2015). Additional lab and theoretical work is urgently needed to understand how such reflective and abundant aerosols are formed in sub-Neptune environments. Some possibilities are a more reflective type of hydrocarbon or darker particles coated in high-albedo condensates (e.g. Lavvas et al. 2019). Upcoming JWST observations should shed light on whether sub-Neptune aerosols share a universal set of properties and whether transitions from clear to aerosol-enshrouded conditions occur at expected levels of insolation.

## 4. ATMOSPHERIC MASS LOSS

Transiting exoplanets are particularly vulnerable to atmospheric mass loss as a result of their close-in orbits. The smallest and most highly irradiated exoplanets may lose their entire atmosphere (see Section 1.2), while the population of close-in gas giant planets appears to be minimally altered by atmospheric mass loss (e.g. Vissapragada et al. 2022; Lampón et al. 2023). For giant planets, atmospheric outflows are driven by high energy irradiation from the host star, which causes the uppermost layers of the planet's atmosphere to expand until they become unbound; this process is often referred to as 'photoevaporation' (for a comprehensive review of theoretical work on this topic, see Owen 2019). For smaller planets, core-powered mass loss (Ginzburg et al. 2018) might also play an important role. For a more detailed discussion of current constraints on these processes from the measured radius-period distribution of sub-Neptune-sized exoplanets, see Rogers et al. (2021) and Owen et al. (2023).

We can directly observe the atmospheric outflows of transiting planets by measuring the depth of the transit in strong atomic absorption lines. The large planet-star radius ratios of transiting gas giant exoplanets make them particularly favorable targets for these observations. The first atmospheric outflows from close-in gas giant planets were detected by measuring the strength of the absorption in the Lyman $\alpha$ line of hydrogen (Vidal-Madjar et al. 2003; Lecavelier Des Etangs et al. 2010). This line is located in the UV, and can therefore only be accessed by space telescopes like HST. Because the core of this line is masked by absorption from the interstellar medium, these observations are only sensitive to absorption in the line wings and are limited to relatively nearby (distances of $\sim$ 100 pc or less) stars. This absorption corresponds to the higher velocity components of the outflow, which are located farther from the planet (e.g., Owen et al. 2023). To date, there are seven planets whose outflows have been measured in this line; see Fig. 11 for a visualization of where these planets are located in mass-period space.

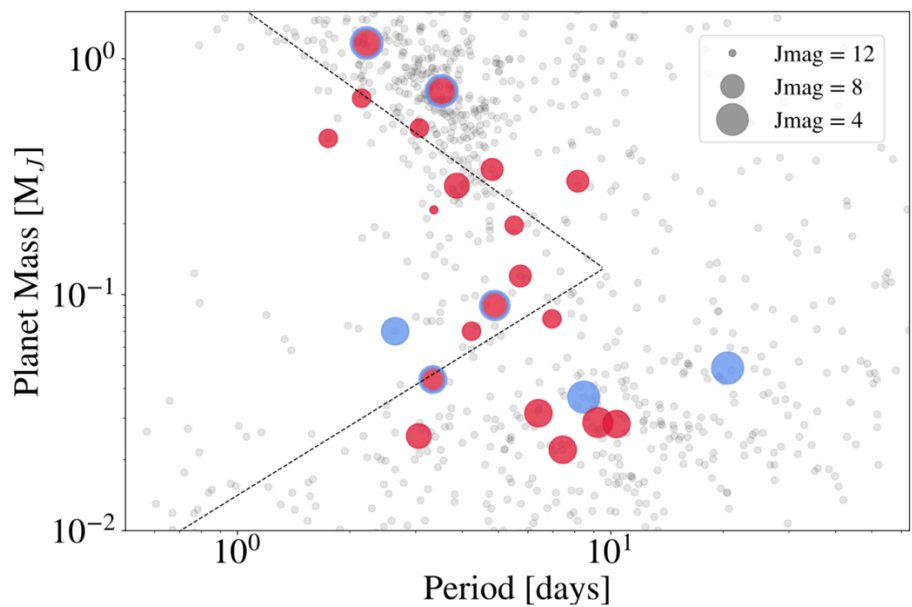

**Figure 11.** Orbital period versus mass distribution for the current sample of planets with measured mass loss rates ($> 3\sigma$ significance; for a complete list see review by Dos Santos 2022) using either Lyman $\alpha$ (blue filled circles; we also include a detection of AU Micb by Rockcliffe et al. 2023), metastable helium (red filled circles, we also include detections for HAT-P-67b, TOI-1268b, TOI-1420b, and TOI-2134b from GullySantiago et al. 2023; P´erez Gonz´alez et al. 2023; Zhang et al. 2023a), or both (blue circles with red fill). The size of the points is scaled according to the host star's brightness in $J$ band (infrared), which depends on the star's distance and mass; brighter stars (smaller $J$ magnitudes) are generally located closer to the Earth and/or have larger masses. The full sample of confirmed planets is shown as grey circles for comparison. There is a deficit of sub-Saturn-sized planets on close-in orbits; this region is called the 'Neptune desert', and its approximate boundaries as defined in Mazeh et al. (2016) are shown as black dashed lines. *Figure courtesy of M. Saidel.*

Recent theoretical (Oklopčić & Hirata 2018) and observational work (Spake et al. 2018; Nortmann et al. 2018) revealed that atmospheric outflows could also be detected using metastable helium absorption at 1083 nm. Unlike Lyman $\alpha$, this line can be readily observed using high resolution spectrographs on ground-based telescopes. Because we can measure absorption in the line core, this line provides a complementary tool to probe the lower velocity components of the outflow, which are located closer to the planet. To date, atmospheric outflows have been measured for 20 planets using this line (see Fig. 11).

Outflows have also been detected in the optical H$\alpha$, H$\beta$, and H$\gamma$ lines (e.g., Jensen et al. 2012; Yan & Henning 2018; Casasayas-Barris et al. 2019; Wyttenbach et al. 2020), as well as UV lines of other atomic species (e.g., Vidal-Madjar et al. 2004; Sing et al. 2019; Dos Santos et al. 2023). Some of the refractory atomic species detected in optical high spectral resolution data sets (see Section 2.3) likely also probe unbound regions of the atmosphere, but more detailed models are needed in order to interpret these absorption signals (Linssen & Oklopčić 2023). By combining the information from multiple lines together, we can obtain a more detailed picture of the overall structure and thermodynamics of the outflow (Lampón et al. 2021; Yan et al. 2022a; Huang et al. 2023; Linssen & Oklopčić 2023).

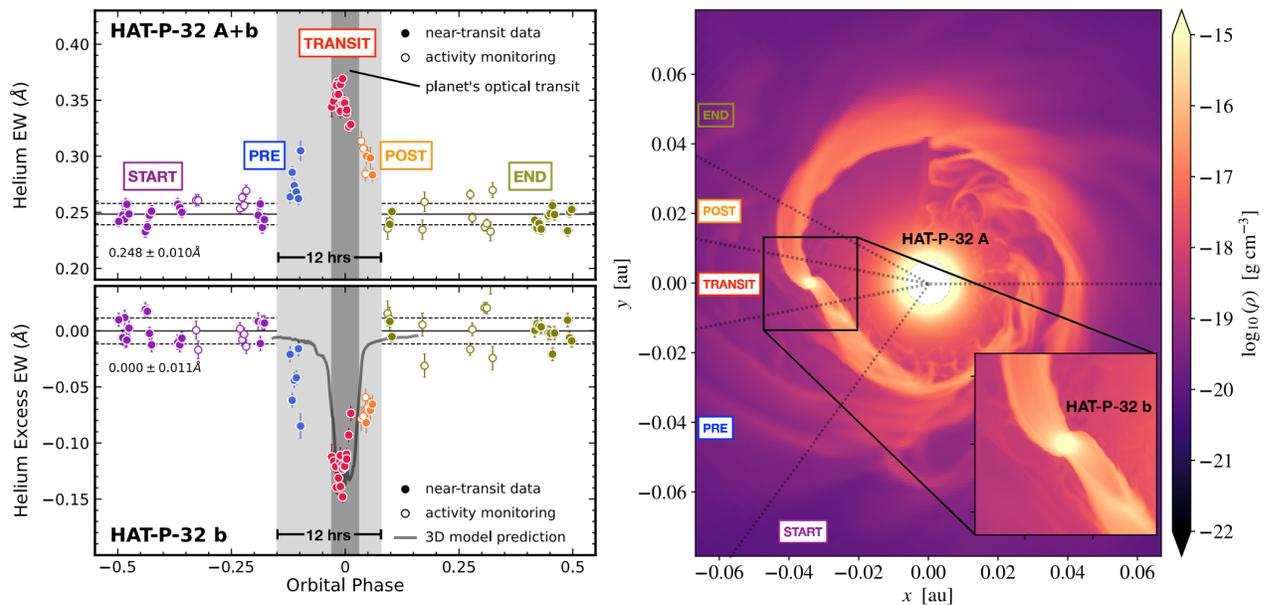

**Figure 12.** Measurement of hot Jupiter HAT-P-32 b's 1083 nm metastable helium absorption signal as a function of orbital phase from Zhang et al. (2023c). The unusually extended nature of this planet's outflow is distinct from that of most other hot Jupiters with published helium detections, which tend to have more narrowly confined outflows. **Left, upper panel:** Helium line equivalent width (EW) for HAT-P-32 b as a function of orbital phase. Solid circles indicate data taken in conjunction with a transit event, while open circles indicate data taken as part of a stellar monitoring program. The phased data are divided into five sections marked by gray shading and colored accordingly. The period where the planet is transiting the star is shown with dark gray shading. Results from a 3D hydrodynamic model are overplotted as a solid gray line. **Left, lower panel:** Equivalent width values after subtracting the average stellar spectrum. **Right:** Slice through the orbital plane of a 3D hydrodynamic simulation of a system with properties similar to that of HAT-P-32. The outflowing gas expands into long tails that lead and trail the planet's orbit, resulting in strong helium absorption before and after the transit. Approximate viewing angles for each colored time segment are shown with colored labels. The logarithmic gas density distribution is indicated using the color bar on the right. *Figure from Zhang et al. (2023c).*

The magnitude of the atmospheric absorption signal during transit can be converted into a mass loss rate by modeling the outflow as a spherically symmetric isothermal Parker wind (Oklopčić & Hirata 2018; Lampón et al. 2020; Dos Santos et al. 2022; Linssen et al. 2022). If the outflow is not spherical but instead is sculpted into a comet-like tail by the stellar wind, we would expect to see an extended absorption signal after the end of the transit egress (e.g., Ehrenreich et al. 2015; Lavie et al. 2017; Kirk et al. 2020; Spake et al. 2021). If there is outflowing material orbiting just ahead of the planet, we may also see absorption prior to the planet's ingress, or even absorption extending many hours before and/or after the planet transit (Zhang et al. 2023c; Gully-Santiago et al. 2023, see Fig. 12). The time-dependent absorption signal, as well as its spectroscopically resolved velocity structure, therefore provide us with important information about the three-dimensional structure of the atmospheric outflow (e.g. Wang & Dai 2021a,b; MacLeod & Oklopčić 2022). In addition to stellar winds, these outflow geometries may also be shaped by the planetary and stellar magnetic field geometries (Owen & Adams 2014; Schreyer et al. 2023; Fossati et al. 2023).

It is more challenging to detect atmospheric outflows from sub-Neptune-sized planets. There are currently only three sub-Neptune-sized planets orbiting mature (> 1 Gyr) stars with published detections (Bourrier et al. 2018; Ninan et al. 2020; Palle et al. 2020b; Orell-Miquel et al. 2022; Zhang et al. 2023a), one of which is disputed (GJ 1214b; see discussion in Spake et al. 2022). Fortunately, these outflows are more easily observable if we focus on the subset of small planets orbiting young stars. Young stars are more active and have enhanced high energy fluxes (e.g., Johnstone et al. 2021; King & Wheatley 2021), while young planets have radii that are still inflated by leftover heat from their formation. As a result, young sub-Neptunes are expected to have enhanced mass loss rates as compared to their more evolved counterparts. Observations of young transiting sub-Neptunes have revealed the presence of atmospheric outflows in both Lyman $\alpha$ (Zhang et al. 2022c) and metastable helium (Zhang et al. 2022a, 2023b; Orell-Miquel et al. 2023). These observations can be used to test the predictions of atmospheric mass loss models seeking to explain the origin of the bimodal radius distribution of small close-in planets (see Section 1.2).

## 5. DAYSIDE TEMPERATURE STRUCTURE

### 5.1. *The Physics of Thermal Inversions*

Measuring the thermal structure of exoplanet atmospheres provides critical insight into how energy is transported and deposited in planetary envelopes. In the solar system, for example, we know that Earth has a stratospheric thermal inversion due to UV/optical absorption by the $O_3$ molecule. Venus has a lower equilibrium temperature than the Earth, despite receiving nearly twice as much energy from the Sun as the Earth does, due to its high Bond albedo. Mercury has a scalding hot dayside and a frigid nightside because it lacks a thick atmosphere to transport heat. All of these types of processes can be assessed by measuring the dayside temperatures and vertical temperature gradients in exoplanet atmospheres.

Thermal inversions in particular have been an interesting phenomenon accessed via dayside thermal emission spectra. Planetary atmospheres that are strongly absorbing at the wavelengths at which their host stars puts out most of their energy will experience heating at the location where the stellar energy is deposited. To ensure global energy balance, this heating comes at a cost of cooling regions deeper in the atmosphere, thus creating a thermal inversion in which temperature *increases* with altitude, peaking around the region where the starlight is absorbed (i.e. the $\tau \sim 1$ surface, where $\tau$ is the optical depth). Spectroscopically, thermal inversions are identified by observing spectral lines in emission, as opposed to absorption lines, which are seen when temperature decreases outwardly. The shape of spectral features relative to the surrounding continuum is therefore used to assess the temperature gradient in the observable portion of an exoplanet atmosphere via thermal emission measurements. By detecting a thermal inversion and simultaneously measuring atmospheric composition, astronomers can also attempt to infer which absorber(s) are responsible for the upper atmosphere heating. As discussed in Section 2.3, TiO, VO, and a variety of refractory species have been proposed as optical and UV absorbers that can generate thermal inversions in hot Jupiters (e.g. Fortney et al. 2008; Lothringer et al. 2018). Other opacity sources such as hazes, clouds, or even water vapor for planets orbiting M-dwarfs have been proposed to similarly drive thermal inversions in cooler planets (e.g. Morley et al. 2015; Arney et al. 2016; Malik et al. 2019; Lavvas & Arfaux 2021; Roman et al. 2021). Because thermal inversions are generated by absorption of incident stellar energy, they are primarily expected to be a dayside phenomenon, although efficient horizontal heat exchange can cause them to persist away from the sub-stellar point and even around to a planet's nightside (e.g. Komacek et al. 2022).

## 5.2. *The Hot-to-Ultrahot Jupiter Transition*

Forward models of hot Jupiter emission spectra have long predicted a transition in dayside thermal structure from planets with inversions to those without, as a function of decreasing planetary temperature. The thermal inversions would be driven by gas-phase optical and UV absorbers that condense out of the atmosphere at lower temperatures, thus rendering the atmosphere more transparent to stellar irradiation (and therefore producing un-inverted temperature profiles) at lower equilibrium temperatures (e.g. Hubeny et al. 2003). Fortney et al. (2008) initially proposed that TiO and VO should be the key drivers of thermal inversions, resulting in a transition to inverted temperature profiles around a planetary equilibrium temperature of 1500 K. Evidence of thermal inversions from secondary eclipse spectra probing planets around this cutoff temperature with Spitzer observations was initially mixed (e.g. Richardson et al. 2007; Charbonneau et al. 2008; Knutson et al. 2008, 2009; Deming et al. 2011; Todorov et al. 2010, 2012, 2013; Baskin et al. 2013; Diamond-Lowe et al. 2014). Ultimately, improved spectroscopic investigations with the HST+WFC3 instrument and high resolution ground-based spectrographs clearly demonstrated un-inverted temperature profiles in various hot Jupiters around and above the predicted 1500 K cutoff temperature, via spectral features appearing in absorption (Birkby et al. 2013; Kreidberg et al. 2014b; Schwarz et al. 2015; Line et al. 2016, 2021). This was accompanied by failures to definitively detect gas-phase TiO and VO in transmission spectra of some of the same planets, implying removal via nightside condensation coldtrapping or some other disequilibrium chemistry process, or perhaps a more mundane explanation such as inaccurate line lists (e.g. D´esert et al. 2008; Hoeijmakers et al. 2015).

Even hotter planets were ultimately required to produce definitive evidence for thermal inversions. 'Ultrahot' Jupiters, as discussed in Section 2.2 are those that are so hot that water dissociates in their atmospheres, and various refractory elements (not just Ti and V) are predicted to be in the gas phase (Parmentier et al. 2018; Lothringer et al. 2018). In these planets, temperature inversions are predicted to be helped along by gas-phase metals and oxides such as Fe, Mg, SiO, etc. Formally the cutoff between hot and ultrahot Jupiters occurs around $T_{eq}$ = 2200 K. The first ultrahot Jupiter to produce a clear detection of a dayside thermal inversion was WASP-121 b (Evans et al. 2017). The 1.4 $\mu m$ water feature in this planet's secondary eclipse spectrum appears in emission, although the feature is quite subtle. Other ultrahot Jupiters, as mentioned in Section 2.2, produced nearly featureless secondary eclipse spectra across the WFC3 bandpass, leading to ambiguous interpretation as to whether water was simply absent from these atmospheres or the dayside temperature profiles were isothermal, thus masking any spectral features (Sheppard et al. 2017; Mansfield et al. 2018; Kreidberg et al. 2018; Mansfield et al. 2021).

The picture of a transition to ultrahot planets with thermal inversions becomes clearer with population-level studies. When looking at WFC3 thermal emission spectra vs. the planets' measured dayside temperatures, Mansfield et al. (2021) identify a clear trend from un-inverted temperature profiles at lower dayside temperatures, to inverted profiles at dayside temperatures above ~2500 K (Figure 13). This is in line with predictions from forward models, although such models still predict the transition to thermal inversions to occur at somewhat lower temperatures. Interestingly, both the models and the data reveal a shift back to featureless spectra with WFC3 at even higher dayside temperatures ($T_{day} \gtrsim$ 3000 K), corresponding to full removal of atmospheric $H_2O$ via thermal dissociation. Another population-level prediction is that ultrahot planets orbiting earlier-type (i.e. hotter) host stars should produce even larger thermal inversions because the peak of the stellar spectral energy distribution (SED) aligns particularly well with the expected UV/optical opacity sources in the planets' atmospheres. This prediction has played out in secondary eclipse observations of the ultrahot Jupiter KELT-20b, which orbits a hot A-type host star. For this planet, the 1.4 $\mu$m $H_2O$ feature appears strongly in emission, much more so than for comparably irradiated planets orbiting later-type G stars (Fu et al. 2022).

Recent JWST and high-resolution emission spectroscopy studies have solidified our understanding of the hot-to-ultrahot Jupiter transition by providing increased precision and wavelength coverage. For example, JWST has the power to resolve the subtle shape of spectral features that were previously hidden in the noise of HST observations. In the case of the the ultrahot Jupiter WASP-18 b, the planet's emission spectrum, which was nearly featureless in HST observations (Sheppard et al. 2017; Arcangeli et al. 2018) is now revealed by JWST to contain very subtle water features in emission, thus confirming the presence of a thermal inversion (Coulombe et al. 2023). In contrast, the cooler 'normal' hot Jupiter HD 149026b shows spectral features in absorption, including clear detections of $H_2O$ and a strong $CO_2$ feature implying high metallicity (Bean et al. 2023). With high-resolution observations from the ground, the emission spectra of various ultrahot Jupiters also provide clear indications of emission lines, demonstrating inverted thermal structures (e.g. Kasper et al. 2021; Yan et al. 2022b, 2023; Brogi et al. 2023; van Sluijs et al. 2023). In at least one case, the detection of a thermal inversion is (finally) accompanied by a high-confidence detection of TiO in the transmission spectrum of the same planet (Yan et al. 2020; Prinoth et al. 2022).

In summary, with newer and better data and population-level studies, astronomers are now finding that the dayside thermal structures of hot and ultrahot Jupiters appear to align with the basic predictions of forward models, albeit with a transition to inverted temperature profiles occurring at somewhat higher equilibrium temperatures than what is predicted for solar composition atmospheres. The forward models assume thermochemical equilibrium and 1-D radiative-convective energy balance, with gas-phase metals and metal oxides serving as strong optical and UV absorbers at high temperatures. Additional new frontiers that will be opened with JWST in the near term include studies of the thermal structures of even colder giant planets. At lower temperatures, dayside clouds or even haze might play a primary role in mediating the deposition of stellar energy in the planets' atmospheres. Hints of such effects already exist in the WFC3 emission spectra of the coolest hot Jupiters investigated to-date (Crouzet et al. 2014; Mansfield et al. 2021). JWST will also enable more detailed studies of the *3-D* structure of giant planet daysides, which will be discussed in more detail in Section 6.

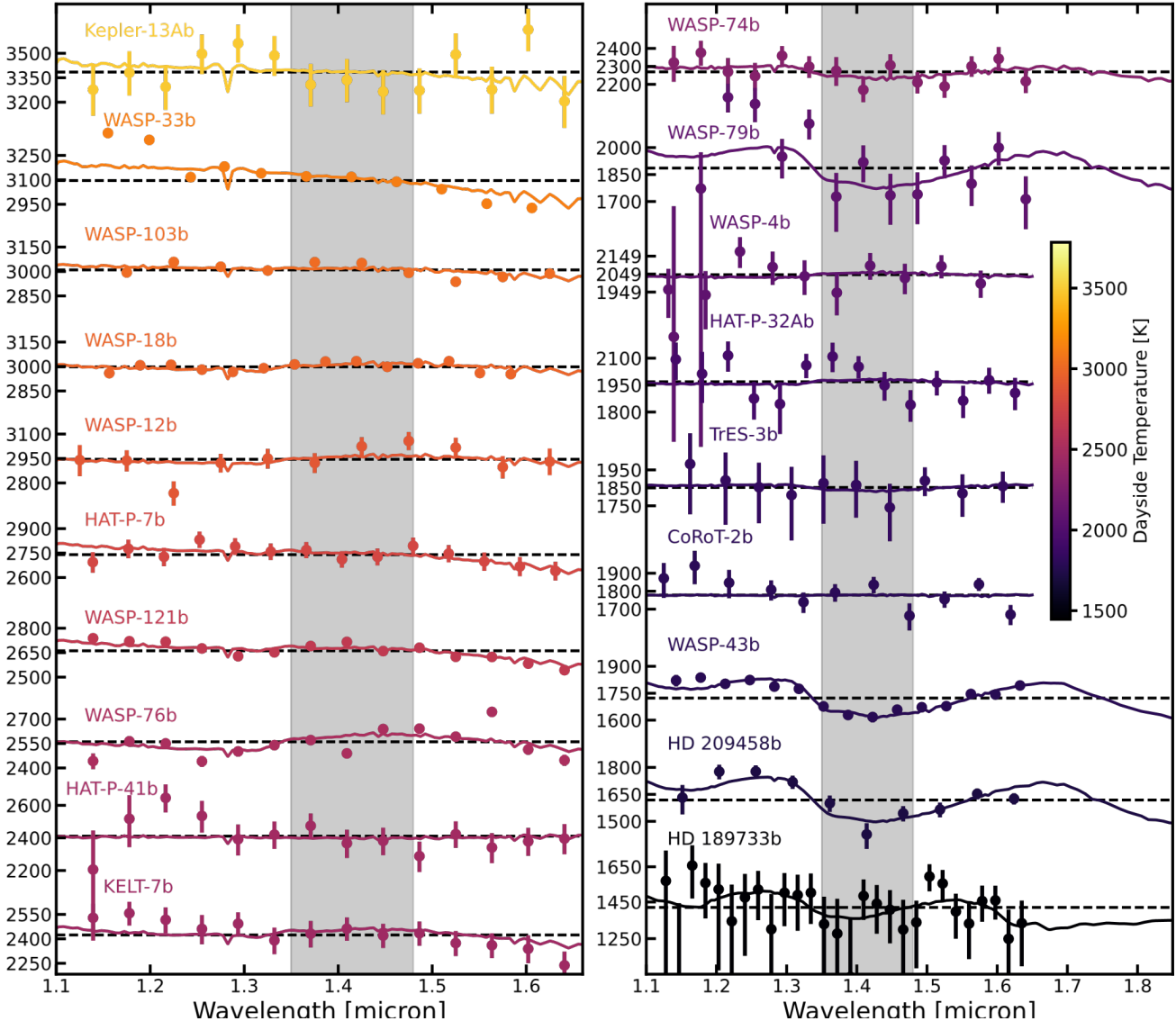

**Figure 13.** Brightness temperature vs. wavelength for hot Jupiter secondary eclipse spectra observed by HST with the WFC3 instrument. Brightness temperature is a measure of the approximate temperature of the photosphere at the wavelength being observed. The 1.4 $\mu m$ water band is indicated by the gray shaded region. For less irradiated planets (toward the bottom right of the plot), the 1.4 $\mu m$ water band appears in absorption, indicating atmospheres with non-inverted temperature profiles. For hotter planets, the absorption features disappear, and in some cases (e.g. WASP-76b, WASP-121b, WASP-12b), the 1.4 $\mu m$ water band subtly inverts into emission, indicating a possible thermal inversion. When compared against forward models of hot Jupiter emission spectra, these observations align well with predictions that thermal inversions occur for planets hotter than ∼2000 K, and water dissociation reduces the abundance of $H_2O$ in the dayside atmosphere. *Figure adapted from Mansfield et al. (2021).*

<center>5.3. *Hot Jupiter Albedos*</center>

Efforts to measure the reflected light from hot Jupiters at optical wavelengths began soon after the discovery of such planets, in order to constrain their albedos. These studies initially resulted in non-detections and upper limits, some of which were quite constraining (e.g. Charbonneau et al. 1999; Rowe et al. 2008; Winn et al. 2008). It was quickly realized that the implied low albedos were in line with the predictions from radiative transfer models for such planets. In the absence of dayside clouds, strong optical absorption lines such as those from Na, K, TiO, etc. absorb out much of the incident stellar radiation, while the only source of reflected light is Rayleigh scattering from the gaseous atmosphere (Seager et al. 2000; Burrows et al. 2008). For cooler planets, in which reflective dayside clouds are expected, geometric albedos should be higher (e.g. Cahoy et al. 2010; Adams et al. 2022), but the general trend of shallower secondary eclipses with lower levels of insolation typically makes it more challenging to detect such signals.

Space telescopes such as CoRoT, Kepler, and TESS were ultimately able to detect the optical secondary eclipses of a number of hot and ultrahot Jupiters, although the broad photometric bandpasses of these facilities has meant that it is typically not possible to fully disentangle the relative contributions of thermal emission vs. scattered light, resulting in model-dependent albedo inferences (e.g. Alonso et al. 2009; Christiansen et al. 2010; Demory et al. 2011). A compilation of optical secondary eclipse measurements for 21 planets with CoRoT, Kepler, and TESS reveals that most such detections have been made at less than 3-$\sigma$ confidence, with inferred geometric albedos ranging between 0 and $\sim$0.3 (Wong et al. 2020). One notable exception is the planet Kepler-7b, which has an inferred albedo of $\sim 0.25$–$0.35$, measured at high confidence (Demory et al. 2011; Wong et al. 2020). This is consistent with the planet's (relatively) low dayside temperature of $\sim$1000 K and the expectation that such conditions are conducive to the formation of reflective clouds. More recently, the European CHEOPS satellite has demonstrated its ability to produce well-constrained measurements of hot Jupiter geometric albedos (Brandeker et al. 2022; Krenn et al. 2023). The inferred values for the hot Jupiters HD 209458b and HD 189733b from CHEOPS lightcurves are 0.096 ± 0.016 and 0.076 ± 0.016. These albedos are far lower than for any solar system planets but in line with models of hot Jupiters having cloud-free dayside atmospheres. In summary, hot Jupiters are dark, but cooler giant planets may be more reflective.

### 5.4. *The Dayside Temperatures of Sub-Neptunes and Super-Earths*

Detecting the thermal emission from smaller and typically cooler sub-Neptunes and superEarths is a much more technically challenging endeavor than for hot Jupiters. Because of this, such studies have mostly been limited to simply detecting a secondary eclipse and measuring an associated brightness temperature, as opposed to full spectroscopic characterization. Once measured, the dayside temperature of the planet can then be used to obtain a combined constraint on both daynight heat redistribution and albedo. All tidally-locked planets have a maximum dayside temperature that can be achieved if the planet's only energy source is the irradiation from its host star:

$$T_{max} = T_* \sqrt{\frac{R_*}{d}} \left(\frac{2}{3}\right)^{1/4}.$$

(5)

This is simply Equation 3 taken in the limit of no day-night heat redistribution (instantaneous reradiation) and zero albedo. Lower measured dayside temperatures are indicative of either a reflective planet or considerable day-night heat transport (or some combination thereof; Koll et al. 2019; Mansfield et al. 2019; Koll 2022).

To date there have only been successful thermal emission detections for two sub-Neptunes: the planets TOI-824b (Roy et al. 2022) and GJ 1214b (Kempton et al. 2023). The former is a hot dense sub-Neptune, whereas the latter is a cooler planet that was already known to have a thick aerosol layer from transmission spectroscopy measurements (see Section 3.3). The dayside temperature of TOI-824b is consistent with its $T_{max}$, whereas GJ 1214b is significantly colder. For sub-Neptunes, a maximally hot dayside, implying poor day-night heat redistribution, requires a high mean molecular weight atmosphere. This result comes from 3-D general circulation models, which demonstrate that heat transport efficiency decreases as a function of increasing mean molecular weight (e.g. Kataria et al. 2014; Charnay et al. 2015; Zhang & Showman 2017). Hydrogen-rich, solar-composition subNeptune atmospheres are predicted to transport heat very efficiently, resulting in cooler daysides and nearly homogeneous global temperatures. Conversely, GJ 1214b's dayside temperature is colder than even its zero-albedo temperature in the limit of fully efficient day-night heat transport, meaning the planet must have a non-zero albedo. This interpretation is confirmed by a full-orbit phase curve with JWST that is best fit by a high mean molecular weight atmosphere coupled with the presence of highly reflective aerosols (see Sections 3.3 and 6).

GJ 1214b is also the only planet smaller than Neptune to have spectral features identified in its dayside thermal emission spectrum. Subtle departures from a blackbody shape imply the presence of geseous water in this planet's atmosphere and a non-inverted temperature profile (Kempton et al. 2023). Interestingly, for planets orbiting M-dwarf host stars, water vapor can actually serve as a source of thermal inversions (Malik et al. 2019; Selsis et al. 2023). This is because its strong near-IR opacity efficiently absorbs stellar light, which in this case peaks at red to near-IR wavelengths. The predicted thermal inversions are fairly weak and high up in the planets' atmospheres though, making their observable consequences negligible for low-resolution spectroscopy with JWST.

Rocky planet thermal emission measurements with Spitzer and more recently JWST have focused on measuring dayside temperatures (as well as phase curves in certain cases) to constrain the presence or absence of an atmosphere. Rocky planets without atmospheres have no mechanism by which to transport heat to their nighsides (Seager & Deming 2009; Koll et al. 2019; Koll 2022). Furthermore, many kinds of rocks that are known to form planetary surfaces in the solar system are very dark[14] (Hu et al. 2012; Mansfield et al. 2019). It therefore can be concluded that a terrestrial planet with a maximally hot dayside temperature is unlikely to have an atmosphere, whereas colder dayside temperatures imply the presence of an atmosphere. Several terrestrial planets to-date have been subjected to this 'secondary eclipse test' to measure their dayside temperatures, with the conclusion in the majority of cases being to rule out the presence of thick atmospheres to varying degrees of confidence (Kreidberg et al. 2019; Crossfield et al. 2022; Whittaker et al. 2022; Greene et al. 2023; Ih et al. 2023, see Section 2.5). The coldest terrestrial planet yet observed in secondary eclipse is TRAPPIST-1c. For that planet, its dayside temperature is only $\sim$2-$\sigma$ consistent with its $T_{max}$ value (Zieba et al. 2023). In this case, the presence of a thick atmosphere is not definitively ruled out, implying that perhaps less irradiated planets are more likely to retain their atmospheres, even if they orbit active M-dwarf stars. Further measurements of rocky planet secondary eclipses with JWST will continue to map out the parameter space of which planets do and do not possess atmospheres, with many such observations already planned for Cycle 2.

---

[14] The assumption of dark planetary surfaces breaks down in the habitable zone and at very small orbital separations, where reflective surfaces are possible (Mansfield et al. 2019).

## 6. THREE-DIMENSIONAL ATMOSPHERIC STRUCTURE

Close-in exoplanets are expected to be tidally locked, with permanent day and night sides. As a result, they can exhibit relatively large day-night temperature gradients, along with corresponding gradients in their atmospheric chemistries and cloud properties. Importantly, tidally locked planets will have relatively slow rotation periods (on the order of days) compared to the gas giant planets in the solar system. This means that the typical length scales for their atmospheric circulation patterns will be much larger ($\sim$ hemisphere-scale) than those of planets like Earth, Jupiter, or Saturn. For a review of the fundamental principles and relevant dynamical regimes for atmospheric circulation on close-in gas giant planets, we recommend Showman et al. (2010) and Showman et al. (2020).

### 6.1. *Fundamentals of Day-Night Heat Transport on Hot Jupiters*

There is a considerable body of observational constraints on the atmospheric circulation patterns of hot Jupiters. During its sixteen years of operation, the Spitzer Space telescope measured broadband infrared secondary eclipse depths for more than a hundred close-in gas giant exoplanets (e.g., Baxter et al. 2020; Wallack et al. 2021; Deming et al. 2023). It also measured broadband infrared phase curves for several dozen gas giant exoplanets (e.g. Bell et al. 2021; May et al. 2022). There are only a few planets with spectroscopic phase curves measured with HST (Stevenson et al. 2014; Kreidberg et al. 2018; Arcangeli et al. 2019, 2021; Mikal-Evans et al. 2022) and (more recently) JWST (MikalEvans et al. 2023b; Kempton et al. 2023; Bell et al. 2023b). Lastly, there are currently two published secondary eclipse maps of the dayside atmospheres of these planets, one from Spitzer (Majeau et al. 2012; de Wit et al. 2012) and one from JWST (Coulombe et al. 2023).

There are several big-picture takeaways that have emerged from the current body of observations. First, both models (Perez-Becker & Showman 2013; Komacek & Showman 2016; Komacek et al. 2017) and observations (e.g., Wallack et al. 2021; Bell et al. 2021; May et al. 2022; Deming et al. 2023) agree that the most highly irradiated gas giant exoplanets have a lower day-night heat redistribution efficiency (defined as the fraction of energy incident on the dayside that is transported to the night side by atmospheric winds) than their more moderately irradiated counterparts. This means that the most highly irradiated gas giant exoplanets have relatively large day-night temperature contrasts, while their cooler counterparts tend to have more uniform temperature distributions (see Figure 14).

These same data also indicate that most close-in gas giant exoplanets have a super-rotating (eastward) equatorial band of wind that transports energy from the day side to the night side, in good agreement with predictions from general circulation models (see Figure 15 and review by Showman et al. 2020). This is readily apparent in infrared Spitzer phase curve observations (Bell et al. 2021; May et al. 2022), which show that the hottest region on the day side is shifted eastward of the substellar point for most hot Jupiters (this corresponds to a phase curve that peaks just before the secondary eclipse). There are several notable exceptions to this trend, which we discuss in more detail later in this section. We can also see the effects of atmospheric circulation in high resolution emission and transmission spectroscopy, where we can directly measure the Doppler shift induced by the planet's atmospheric winds. This can manifest as either a net shift in the lines for a single coherent flow direction, or an overall broadening of the lines for observations that integrate over multiple flow directions (e.g. Miller-Ricci Kempton & Rauscher 2012; Showman et al. 2013; Beltz et al. 2021, 2022). Doppler shifts due to atmospheric winds have been seen in high resolution transmission spectroscopy, which probes the day-night terminator region (e.g. Snellen et al. 2010;

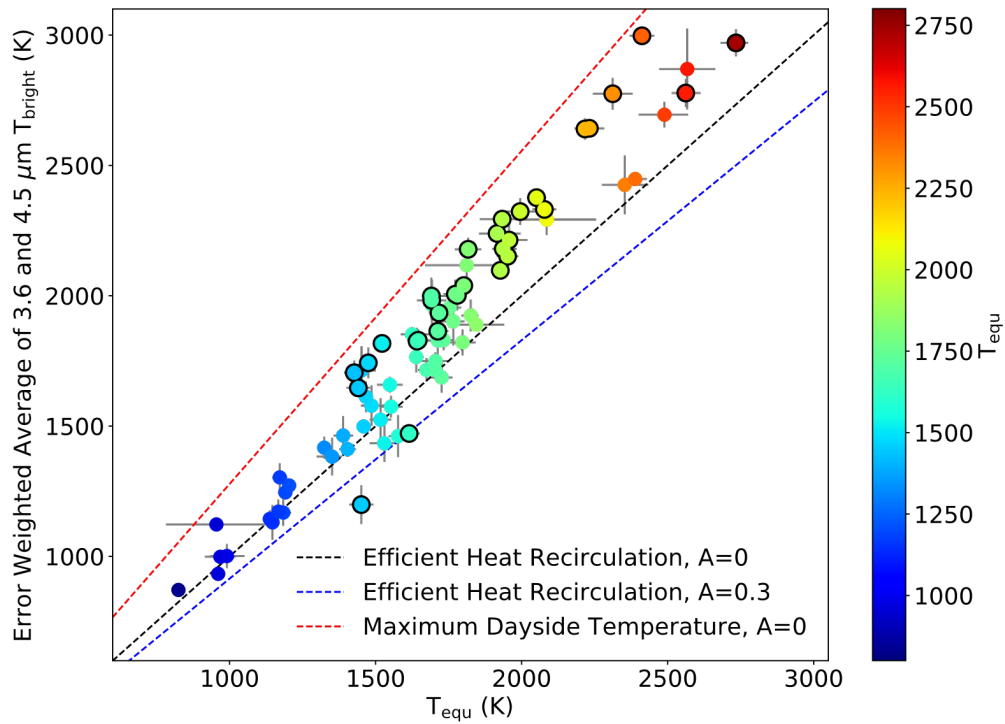

**Figure 14.** Dayside brightness temperatures for the sample of hot Jupiters with measured eclipse depths in the 3.6 and 4.5 μm bands with Spitzer as a function of their predicted equilibrium temperatures. Planets with relatively inefficient day-night recirculation will lie closer to the red dashed line (maximum dayside temperature assuming zero recirculation and zero albedo), while planets with relatively efficient day-night circulation will lie closer to the black dashed line (complete day-night recirculation of energy, zero albedo). The subset of hot Jupiters whose dayside albedos are enhanced by reflective silicate clouds (equilibrium temperatures near 1500 K) can also lie below the black line, as indicated by the blue dashed line. Planets with black circles have spectral slopes that are inconsistent with that of a blackbody across the 3.6 to 4.5 μm band, indicating the presence of strong molecular features. *Figure from Wallack et al. (2021).*

Louden & Wheatley 2015; Brogi et al. 2016; Flowers et al. 2019; Seidel et al. 2021; Kesseli et al. 2022; Pai Asnodkar et al. 2022; Gandhi et al. 2022), and in emission spectroscopy, which integrates over the dayside atmosphere (e.g. Yan et al. 2023; Lesjak et al. 2023).

### 6.2. *Complications from Clouds, Chemical Gradients, and Magnetic Fields*

The non-uniform temperature distributions in the atmospheres of close-in gas giant planets also have important implications for their condensate cloud properties. Clouds that can condense in one region of the atmosphere may not be able to condense in other regions; this can lead to hemisphere-sized cloudy and clear regions in the atmospheres of these planets (Figure 15, and for model predictions of the irradiation-dependent cloud distributions, see Parmentier et al. 2018, 2021; Roman et al. 2021). We can see empirical evidence for patchy clouds in the optical phase curves of close-in planets, which exhibit localized regions of high albedo due to the presence of reflective silicate clouds (Demory et al. 2013). When viewed in transmission, the properties of clouds and/or hazes are also expected to differ between the dawn and dusk terminators. This effect will cause the shape of the transit ingress (when the planet is entering the disk of the star) to differ from that of the transit egress (when the planet is exiting the disk of the star), and should be detectable with JWST (Kempton et al. 2017; Powell et al. 2019; Espinoza & Jones 2021; Steinrueck et al. 2021; Carone et al. 2023). Close-in gas giant planets also appear to have surprisingly uniform nightside temperatures, and it has been suggested that this may be due to the presence of nightside clouds (Keating et al. 2019; Gao & Powell 2021). If confirmed by JWST, this would have important consequences for atmospheric circulation patterns on hot Jupiters, as the presence of these clouds can inhibit radiative cooling on the planet's night side, resulting in a globally hotter atmosphere and a reduced offset for the dayside hot spot (Parmentier et al. 2021; Roman et al. 2021).

These day-night temperature gradients can also lead to chemical gradients between the dayside and nightside atmospheres. For the most highly irradiated hot Jupiters, current observations suggest that refractory species may condense on the night side, even when the dayside atmosphere is hot enough for them to remain in gas phase (Lothringer et al. 2022; Pelletier et al. 2023). High resolution transmsission spectroscopy has also been used to argue for gradients in composition between the dawn and dusk terminators (Ehrenreich et al. 2020; Mikal-Evans et al. 2022; Prinoth et al. 2022, 2023; Gandhi et al. 2022). These chemical gradients can complicate efforts to measure wind speeds using high resolution spectroscopy (e.g. Wardenier et al. 2023; Savel et al. 2023). Recent studies have also explored the role that the dissociation of $H_2$ on the day side and its subsequent recombination on the night side might play in day-night energy transport in these highly irradiated atmospheres (Bell & Cowan 2018; Tan & Komacek 2019; Mansfield et al. 2020; Roth et al. 2021; Changeat 2022).

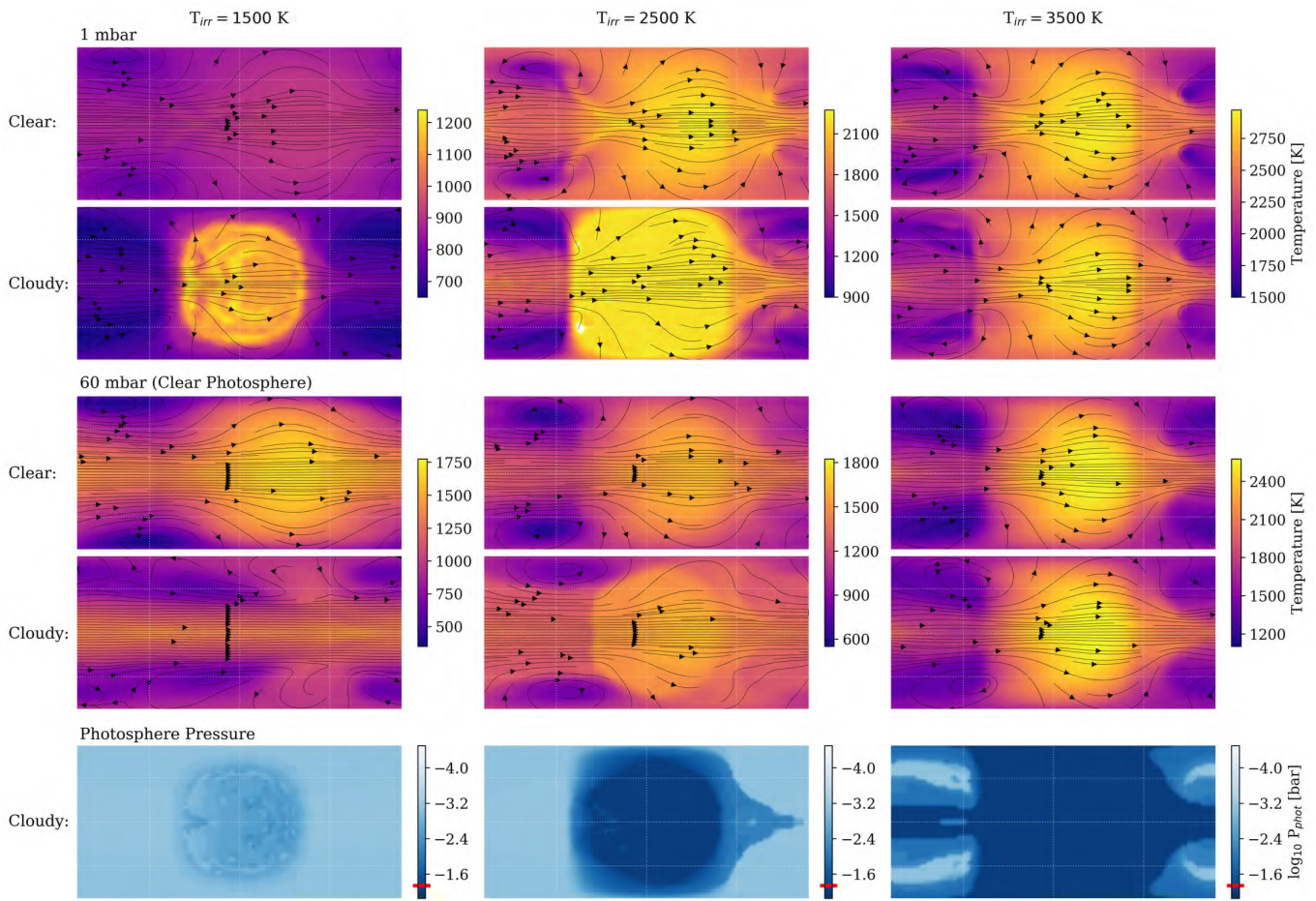

**Figure 15.** Properties of hot Jupiters with three different incident flux levels from general circulation models with and without radiatively active clouds included. The top two rows show the temperature distribution at the top of the atmosphere (1 mbar), and the middle two rows show the temperature distribution slightly deeper down, at the approximate level of the infrared photosphere. The bottom panel shows how the presence of clouds alters the level of the photosphere by increasing the atmospheric opacity; more cloudy regions have lower photospheric pressures, meaning that we do not see as deep into these cloudy regions. The approximate photospheric pressure for the clear atmosphere (60 mbar) is indicated by a red line on the color bar. *Figure from Roman et al. (2021).*

There is also emerging evidence suggesting that atmospheric flow patterns on the most highly irradiated hot Jupiters may be altered by magnetic effects. At these temperatures, the atmosphere consists of a mixture of neutral and ionized species. If the planet has a strong magnetic field, this can lead to magnetically induced drag and correspondingly weakened day-night energy transport (Perna et al. 2010; Menou 2012). Observationally, this would have the effect of moving the hot spot on the day side closer to the substellar point. Recent observations of the ultra-hot Jupiter WASP-18 b with JWST indicate that its relatively small dayside hot spot offset is best matched by circulation models with enhanced drag due to MHD effects (Coulombe et al. 2023). Other ultra-hot Jupiters also appear to have similarly small hot spot offsets in their Spitzer phase curves (Bell et al. 2021; May et al. 2022). As the magnetic field increasingly dominates the atmospheric flow patterns, it may cause the location of the dayside hot spot to vary from orbit to orbit, perhaps even shifting it westward of the substellar point (i.e. opposite of the predicted wind direction for a neutral atmosphere; Rogers 2017; Hindle et al. 2021a,b). This may explain the westward and/or time-varying hot spot offsets of several hot Jupiters (e.g. Dang et al. 2018; Bell et al. 2019). Several planets also appear to have time-varying optical phase curves (Armstrong et al. 2016; Jackson et al. 2019a,b), although in some cases these variations may be the result of stellar and/or instrumental variability (Lally & Vanderburg 2022).

### 6.3. *Circulation Patterns of Sub-Neptune-Sized Planets*

There are currently only a few sub-Neptune-sized planets with phase curve observations. As discussed in Section 5.4, for rocky planets these phase curves can be used to infer the presence or absence of a thick atmosphere based on the observed day-night temperature gradient (Seager & Deming 2009). For planets with thick atmospheres, the shape of the phase curve can also be used to constrain the planet's atmospheric composition. A recent JWST observation of the midIR phase curve of the sub-Neptune GJ 1214 b by Kempton et al. (2023) indicates that this planet likely possesses a high mean molecular weight atmosphere with highly reflective clouds or hazes. Spitzer phase curves of hot rocky super-Earths K2-141 b and LHS 3844 b indicate that these planets have large day-night temperature gradients, suggesting that their atmospheres must be relatively tenuous, if they possesses one at all (Kreidberg et al. 2019; Zieba et al. 2022). Although the Spitzer IR phase curve of the super-Earth 55 Cnc e initially appeared to require the presence of a thick atmosphere (Demory et al. 2016b), a subsequent re-analysis of these data resulted in a larger daynight temperature gradient more in line with those observed for other hot rocky super-Earths (Mercier et al. 2022). Puzzlingly, this planet also appears to have a time-varying infrared flux from its dayside (Demory et al. 2016a), along with a variable optical phase curve (Meier Vald´es et al. 2023). This may indicate the presence of a tenuous, time-varying outgassed high mean molecular weight atmosphere (Heng 2023). JWST will soon observe phase curves for multiple additional rocky exoplanets and sub-Neptunes, expanding our understanding of their atmospheric properties.

## 7. CONCLUSIONS

With the launch of JWST and the advent of high-resolution spectrographs on large groundbased telescopes, the exoplanet atmospheres field has entered a new era. As detailed in the previous sections, we are now reliably measuring chemical abundances and abundance ratios, global temperature fields, wind speeds, and atmospheric escape rates. At the time of writing this article we are just over one year into the JWST mission, and we are already seeing results that are fundamentally shifting our understanding of exoplanet atmospheres. Some early takeaways include the diversity of chemical inventories in giant planet atmospheres and an apparent lack of atmospheres on at least some rocky planets orbiting M-dwarfs. At the same time, a multitude of results from ground-based high-resolution spectroscopy are revealing the richness of (ultra)hot Jupiter chemistry.

With these new observational results come new scientific questions. It is already clear that at this improved level of measurement precision, the 3-D nature of exoplanet atmospheres will need to be carefully taken into account to not bias any scientific conclusions. This challenge is accompanied by new opportunities to directly infer properties of 3-D circulation and weather in exoplanet atmospheres. Measurements of individual exoplanets' spectra are also telling a story about how those planets formed and evolved, but backing out the correct narrative is a truly challenging endeavor, which can be helped along somewhat by population-level investigations. The characterization of smaller exoplanets is one of the key promises of the JWST mission, but new questions have arisen about what subset of such planets even host atmospheres at all and how to disentangle the signatures of stellar activity from atmospheric absorption. Investigations of sub-Neptunes aimed at distinguishing those with primordial atmospheres from a potential population of water worlds must still contend with the confounding influence of aerosols on spectroscopic observations. Along the way, the properties of the aerosols themselves are presenting their own surprises.

The pace of new observational results from JWST and ground-based high-resolution studies is only accelerating. We are in a regime in which our state of knowledge of exoplanet atmospheres in each successive year expands substantially. As such, this review article serves as a snapshot in time of the state of exoplanet observations following the first year of JWST science. We anticipate that some of the open questions presented in this article will be resolved in the near term, whereas others will take future generations of telescopes and scientists to fully answer. What we can surely say is that exoplanet atmospheres have yet to reveal all of their surprises to us.

E.M.R.K. and H.A.K. would like to acknowledge Jacob Bean, Yoni Brande, Thayne Currie, Peter Gao, Jegug Ih, Megan Mansfield, James Rogers, Michael Roman, Morgan Saidel, Arjun Savel, , Nicole Wallack, and Zhoujian Zhang who graciously contributed figures to this review. H.A.K. is also grateful to the Woods Hole Geophysical Fluid Dynamics Program, which provided a thoughtful and interactive venue for developing ideas incorporated into several sections of this review. E.M.R.K. would like to thank Jacob Bean and Tad Komacek for insightful discussions while developing this manuscript.